\documentclass[journal, draftclsnofoot, onecolumn]{IEEEtran}
\usepackage{amsmath,amssymb,amsfonts}
\usepackage{algorithmic}
\usepackage{algorithm}
\usepackage{array}
\usepackage{textcomp}
\usepackage{stfloats}
\usepackage{url}
\usepackage{verbatim}
\usepackage{graphicx}
\usepackage{cite}
\usepackage{mathtools}
\usepackage{siunitx}
\usepackage{xcolor}
\def\BibTeX{{\rm B\kern-.05em{\sc i\kern-.025em b}\kern-.08em
    T\kern-.1667em\lower.7ex\hbox{E}\kern-.125emX}}
\usepackage{bm}
\usepackage{svg}
\usepackage{support-caption}
\usepackage{subcaption}
\usepackage{amsthm}

\newtheorem{remark}{Remark}

\begin{document}

\title{Optimal Pilot Design for OTFS in Linear Time-Varying Channels}
% \author{Ids van der Werf, 
%         Richard Heusdens,
%         Richard C. Hendriks and
%         Geert Leus
%         % <-this % stops a space
% %\thanks{Manuscript received \today.}
% %\thanks{This work has been submitted to the IEEE for possible publication. Copyright may be transferred without notice, after which this version may no longer be accessible.}
% \thanks{This work was partly funded by the Netherlands Organisation for Applied Scientific Research (TNO) and the Netherlands Defence Academy (NLDA), reference no. TNO-10026587.}
% \thanks{All authors are affiliated with the Microelectronics department, Delft University of Technology, Delft, The Netherlands (email: \{i.vanderwerf, r.heusdens, r.c.hendriks, g.j.t.leus\}@tudelft.nl).}
% }

\author{
\IEEEauthorblockN{
    Ids van der Werf\IEEEauthorrefmark{1},
    Richard Heusdens\IEEEauthorrefmark{1}\IEEEauthorrefmark{2},  
    Richard C. Hendriks\IEEEauthorrefmark{1},
    and 
    Geert Leus\IEEEauthorrefmark{1}
    } 
    \\
\IEEEauthorblockA{  \IEEEauthorrefmark{1}Delft University of Technology, Mekelweg 4, 2628 CD Delft, The Netherlands \\ e-mail: \{i.vanderwerf, r.heusdens, r.c.hendriks, g.j.t.leus\}@tudelft.nl, \\ 
                    \IEEEauthorrefmark{2}Netherlands Defence Academy (NLDA), Het Nieuwe Diep 8, 1781 AC Den Helder, The Netherlands \\ e-mail: r.heusdens@mindef.nl}

%
% \thanks{This work was partly funded by the Netherlands Organisation for Applied Scientific Research (TNO) and the Netherlands Defence Academy (NLDA), reference no. TNO-10026587.}
}

\maketitle
%\tableofcontents

\begin{abstract}%
This paper investigates the positioning of the pilot symbols, as well as the power distribution between the pilot and the communication symbols in the orthogonal time frequency space (OTFS) modulation scheme. We analyze the pilot placements that minimize the mean squared error (MSE) in estimating the channel taps. This allows us to identify two new pilot allocations for OTFS. In addition, we optimize the average channel capacity by adjusting the power balance. We show that this leads to a significant increase in average capacity. The results provide valuable guidance for designing the OTFS parameters to achieve maximum capacity. Numerical simulations are performed to validate the findings.
\end{abstract}


\section{Introduction}
\noindent \noindent\IEEEPARstart{T}{o} address the growing need for data, it is important to judiciously consider the design of the modulation scheme. In recent wireless communication standards, orthogonal frequency division multiplexing (OFDM) has been widely adopted as the preferred choice \cite{wang2000wireless}. However, one drawback of OFDM is its vulnerability to Doppler effects in the channel. As communication scenarios increasingly involve dynamic environments, 
%such as applications in traffic \textcolor{red}{[REF]} or underwater communications \textcolor{red}{[REF]}, 
there have been proposals for new modulation schemes that offer improved resistance to Doppler effects. Lately, orthogonal time frequency space (OTFS)~\cite{hadani2016OTFS,hadani2017orthogonal} modulation has received a lot of attention. OTFS defines symbols in the delay-Doppler domain and then transforms the signal into the time domain using the Zak transform~\cite{hadani2018orthogonal}. OTFS has been shown to have improved performance compared to OFDM~\cite{ebihara2014underwater,hadani2017orthogonal,hadani2018orthogonal,raviteja2018interference}, which is attributed to the fact that OTFS can benefit from diversity in time and frequency. Throughout this paper, we will use the name OTFS; however, it is worth mentioning that the (older) modulation schemes vector OFDM (V-OFDM)~\cite{xia2001precoded}, asymmetric OFDM (A-OFDM)~\cite{zhang2007asymmetric} and orthogonal signal-division multiplexing (OSDM)~\cite{ebihara2014underwater} were shown to be equivalent to OTFS~\cite{raviteja2019OTFS,xia2022comments,werf2024onthe}. Therefore, our analysis and conclusions also apply to these modulation schemes.

To get the most out of the OTFS modulation, careful design of the pilot symbols is required. Although many pilot allocations have been proposed for OTFS (see, e.g., \cite{raviteja2018embedded,raviteja2019embedded,qu2021lowdimensional,sanoopkumar2023apractical}), a comparison and (mathematical) analysis of the optimality of these allocations is lacking.
\IEEEpubidadjcol
Therefore, the main contributions of this paper are the following:
\begin{itemize}
   \item We give an overview of the work on optimal pilot design for LTI and LTV channels, and of the work on pilot design for the related modulation schemes (Section \ref{sec:relatedwork}).
   \item We reformulate the effect of the LTV channel on the OTFS modulation (Section \ref{sec:OTFSOSDMproperties}). This allows us to show that the allocations with the lowest pilot overhead achieve the minimum mean squared error (MSE) on the estimation of the channel taps (Sections \ref{subsec:lowestoverhead} and \ref{subsec:MMSEoptimality}). 
   \item We show that the average channel capacity can be significantly increased by choosing the OTFS parameters carefully and by optimizing the power balance between the pilot and the communication symbols. This optimal power balance also drastically decreases the bit error rate (BER).
   \item Finally, our findings can be used as guidance for designing the OTFS parameters to increase the channel capacity.
\end{itemize}

\paragraph*{Notation} 
In what follows, $\otimes$, $\odot$, $\circ$ and $\ast$ are used to denote the Kronecker product, the Khatri-Rao product, the element-wise multiplication and the linear convolution, respectively. Let $K$ be a positive integer and let $\mathbf{P}_K$ denote a $K\times K$ cyclic permutation matrix given by,
\begin{equation*}
    \mathbf{P}_K = 
    \begin{bmatrix}
        0 & 0 & \dots & 0 & 1 \\
        1 & 0 & \dots & 0 & 0 \\
        0 & 1 & 0 &  & \vdots \\
        \vdots & \ddots &\ddots & \ddots& \\
        0 & \dots & 0 & 1 & 0 \\        
    \end{bmatrix},
\end{equation*}
and let $\mathbf{P}^l_K$ denote the $l$'th power of $\mathbf{P}_K$ for some integer $l$. Let $N$ and $M$ be positive integers. If $K = NM$, we can rewrite the cyclic permutation matrix as $\mathbf{P}_K = \mathbf{I}_N \otimes \mathbf{L}_M + \mathbf{P}_N \otimes \mathbf{U}_M$, where $\mathbf{L}_M$ is a cyclic permutation matrix of size $M$ except for the top right element, which is zero, and $\mathbf{U}_M$ is a zero matrix of size $M$ except for the top right element, which is one (thus $\mathbf{P}_M = \mathbf{U}_M + \mathbf{L}_M$). Similarly, we can write $\mathbf{P}^l_K = \mathbf{I}_N \otimes \mathbf{L}^{(l)}_M + \mathbf{P}_N \otimes \mathbf{U}^{(l)}_M$, where the matrices $\mathbf{L}^{(l)}_M$ and $\mathbf{U}^{(l)}_M$ denote the lower and upper part of $\mathbf{P}^l_M$, respectively.
Let $Q$ be an even integer and let $q$ be an integer between $-Q/2$ and $Q/2$, then the matrix $\bm{\Lambda}^{(q)}_K = \textrm{diag}(e^{j 2\pi q \:0/K}, \dots, e^{j 2\pi q (K-1)/K})$.

Before we elaborate on the related work, we start by a brief discussion of the channel in the next section.

\section{Signal model}
\noindent In this section, we discuss the OTFS modulation scheme and the linear time-varying channel model.

\subsection{OTFS transmitter}
\begin{figure}
    \centering
    \includegraphics[width=0.5\textwidth]{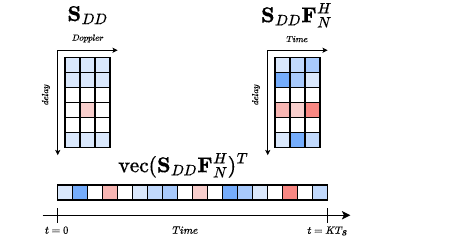}
    \caption{Visualization of the OTFS transmitter, $\{K,N,M\} = \{18,3,6\}$.}
    \label{fig:OTFSmodvis}
    \vspace{-0.5cm}
\end{figure}

\noindent For OTFS, the transmitter characterizes $K = NM$ symbols in the delay-Doppler (DD) domain ($M$ by $N$ symbols in the delay and Doppler dimension respectively) after which the symbols are consecutively transformed to the time-frequency (TF) and the time domain by the inverse symplectic finite Fourier transform (ISFFT) and the Heisenberg transform~\cite{hadani2017orthogonal}, respectively. Assuming a rectangular transmit pulse is used~\cite{raviteja2019practical}, the transmitted signal in discrete-time baseband is given by~\cite{raviteja2019OTFS,werf2024onthe},
\begin{equation}
    \mathbf{x} = \textrm{vec}(\mathbf{S}_\textrm{DD}\mathbf{F}^H_N) 
    %= (\mathbf{F}^H_N \otimes \mathbf{I}_M)\textrm{vec}(\mathbf{S}_\textrm{DD}) 
    = (\mathbf{F}^H_N \otimes \mathbf{I}_M)\mathbf{s},
    \label{eq:OTFStranmitter}
\end{equation}
where $\mathbf{s} = \textrm{vec}(\mathbf{S}_\textrm{DD})$. In Fig. \ref{fig:OTFSmodvis}, a visual representation of the operation at the transmitter, i.e. \eqref{eq:OTFStranmitter}, is shown.

\subsection{Channel Model}\label{sec:channelmodel}
\noindent Let $r(t)$, $h(t,\tau)$, $x(t)$ and $n(t)$ denote the received signal, the channel impulse response, the transmitted signal and the additive channel noise, respectively, then the linear time-variant (LTV) channel can be described by
\begin{equation*}
    r(t) = \int_0^{\infty} h(t,\tau)x(t-\tau)\textrm{d}\tau + n(t).
\end{equation*}
Note that a linear time-invariant (LTI) channel is subsumed by this model; if $h(t,\tau) = h(\tau)$, the model is time-invariant.

Given that we have Nyquist-rate sampling, we can express the discrete-time version using the notation $t_k = kT_s$ and $\tau_l = t_l =  lT_s$. To simplify the notation, we will consider the discrete-time function instead of the continuous-time ones sampled at discrete instances, i.e.
\begin{equation*}
    r(k) = \sum_{l=0}^{L} h(k,l)x(k-l) +n(k).
\end{equation*}
Suppose we collect a total of $K$ samples over time, then estimating all the $K(L+1)$ channel coefficients is an ill-posed problem, as in general $K<K(L+1)$. In order to decrease the number of coefficients that need to be estimated, the basis expansion model (BEM) was introduced~\cite{tsatsanis1996modelling}.

The BEM is a model that represents how the channel changes over time. It approximates the channel taps by expressing them using a lower order basis,
\begin{equation}
    h(k,l) = \sum_{q=-Q/2}^{Q/2} c_{l,q} b_q(k).
    \label{eq:BEM}
\end{equation}
Here $b_q(k)\in\mathbb{C}$ is the function representing the basis.
In general $Q+1\ll K$, and thus the total number of coefficients to be estimated is now lower, casting the problem well posed if $K\geq (Q+1)(L+1)$. 

This raises the question of what basis and thus what function $b_q(k)$, should one use. Over the past decades, many different bases have been proposed. The most well-known is the complex exponential BEM (CE-BEM)\cite{tsatsanis1996modelling}, for which $b_q(k) = e^{j \omega_q k}$, $\omega_q = 2\pi q/K$. Note that the CE-BEM is quite comprehensible; the coefficients of the CE-BEM each represent a unique pair of Doppler shift $\omega_q$ and time delay $\tau_l$. 
Some other popular BEMs are the generalized CE-BEM (GCE-BEM) \cite{leus2004on}, for which $b_q(k) = e^{j \omega_q k}$, $\omega_q = 2\pi q/(KR)$, $R\geq1$, the polynomial BEM (P-BEM) \cite{borah1999frequency}, the discrete Karhuen–Loève BEM (DKL-BEM)~\cite{teo2005optimal} and the discrete prolate spheroidal BEM (DPS-BEM)~\cite{zemen2005timevariant}.
Of course, the modeling accuracy of each choice differs per application.

In the OTFS literature the so-called delay-Doppler channel instead of a BEM is used. However, assuming that the time delays and Doppler shifts fall on the Nyquist grid, the delay-Doppler channel coincides with the (conventional) CE-BEM, as shown in Appendix \ref{app:ddchanisBEM}. 

Assuming that we model the channel with a CE-BEM with additive Gaussian noise with zero mean and covariance $\mathbf{R}_\mathbf{n}$, the received signal is given by
\begin{equation*}
    \mathbf{r} = \mathbf{H}\mathbf{x} + \mathbf{n} = \left( \sum_{q=-Q/2}^{Q/2} \sum_{l=0}^L c_{l,q} \bm{\Lambda}^{(q)}_K \mathbf{P}_K^l \right)\mathbf{x} +\mathbf{n},
\end{equation*}
where the vectors $\mathbf{r}$, $\mathbf{x}$ and $\mathbf{n}$ collect the samples of $r(k)$, $x(k)$ and $n(k)$, respectively, for $k = 0,1,\dots,K-1$.
\begin{remark}\label{rem:CEBEMisDDchan}
    Note that, when a channel is created with the delay-Doppler channel model, usually each delay tap $\tau_i$ (often called a path) has only one Doppler shift $\omega_i$. Note that two paths could be made with the same delay, $\tau_i = \tau_j$, for which $\omega_i \neq \omega_j$, so that a path with a Doppler \textit{spread} is created. This is often forgotten in the literature, e.g. \cite{raviteja2018embedded,shen2019channel,jing2021two}. 
    The false assumption that only one Doppler shift could be used was a reason for \cite{liu2022nearoptimal} to ``switch" to the CE-BEM. Upon noting their equivalence, it is clear that by viewing the delay-Doppler channel model as the CE-BEM, it is more comprehensible that one can create paths with multiple Doppler shifts, i.e. Doppler spread (for every $\tau_l$ one has multiple $\nu_q$, such that one can set $c_{l,q_1}$ and $c_{l,q_2}$ non-zero).
\end{remark}
\begin{remark}
    The parameters $L$ and $Q$ denote the maximum time delay and Doppler shift in the channel. Although specific channel taps vary, these parameters remain constant over a long time. Upper bounds for these values are typically well-known for particular environments, allowing $L$ and $Q$ to be set to these bounds before communication.
\end{remark}

\subsection{OTFS receiver}
\noindent The literature on OTFS receivers is rich; see, e.g. \cite{raviteja2018interference,murali2018on,jing2021two,liu2022nearoptimal}. However, the most straightforward way is apply a minimum mean square error (MMSE) estimator on the demodulated signal, which is given by
\begin{equation*}
    \mathbf{y} = (\mathbf{F}_N \otimes \mathbf{I}_M)\mathbf{r}.
\end{equation*}
To obtain the signal received in the DD domain, the vector $\mathbf{y}$ is reshaped into an $M\times N$ matrix, i.e., $\mathbf{Y} = \textrm{vec}^{-1}(\mathbf{y})$.

\section{Related work on pilot design}\label{sec:relatedwork}  
\noindent In this section, we examine the existing literature on the development of (optimal) pilot symbols. We identify four main research areas that concentrate on pilot design: studies centered on an LTI channel, on an LTV channel, on OTFS modulation, and on OSDM.
\subsection{Optimal pilot design in an LTI channel}
\noindent The investigation of pilot allocation started in the late 1990s. Various performance measures such as the MSE, capacity, Cramér-Rao lower bound (CRB), etc., were taken into account, and it was found that the optimal allocations are quite similar. We will review some of the early studies. 
Note that to obtain an LTI channel from the expression in \eqref{eq:BEM}, one should set $Q = 0$. Moreover, since there is no time-varying component in an LTI channel, one does not need a BEM to model the time variation. Typically, the number of unknowns is already lower than the number of knowns, that is, typically $K>L+1$.

The work in \cite{adireddy2002optimal} shows an (capacity) optimal allocation using single carrier modulation (SCM). They cluster at least $2L+1$  pilot symbols, where the leading and trailing $L$ symbols are set to zero. Setting the leading and trailing $L$ symbols to zero was also found to be optimal in \cite{dong2002optimal}, which uses the CRB on the channel tap estimator as a performance measure. 
The findings are consistent with the fact that the MSE and the CRB coincide when considering a linear model in the presence of additive white Gaussian noise (AWGN).

Using OFDM for transmission, it has been observed that the optimal approach in terms of MSE on the channel taps $\{c_l\}_{l=0}^{L}$~\cite{negi1995pilot} and in terms of (a lower bound on) channel capacity~\cite{adireddy2002optimal} is to modulate pilot symbols on $L+1$ equally spaced frequencies.
Although in \cite{negi1995pilot} and \cite{adireddy2002optimal} it was assumed that the pilot symbols which are modulated on the frequencies, have equal energy, in \cite{ohno2002optimal} it was proved that the equipowered pilot symbols are indeed optimal in terms of MSE on channel taps. Thus, we can conclude that the use of $L+1$ equipowered pilot symbols on equispaced frequencies is the optimal pilot design. 
Moreover, the (capacity) optimal energy distribution between the pilot symbols and the communication symbols was derived in closed form \cite{adireddy2002optimal,ohno2002optimal}.
In \cite{ohno2004capacity} the authors extend the work in \cite{ohno2002optimal} by considering a probabilistic/stochastic channel instead of a deterministic one, and equivalent conclusions are drawn.

\subsection{Optimal pilot design in an LTV channel}
\noindent In this section, we discuss the most important work on optimal pilot design while assuming an LTV channel. Unless pointed out differently, the works model the LTV channel with a CE-BEM.

\begin{figure*}[htbp]
\centering
\begin{subfigure}{.33\textwidth}
  \centering
    \includegraphics[width=1\textwidth]{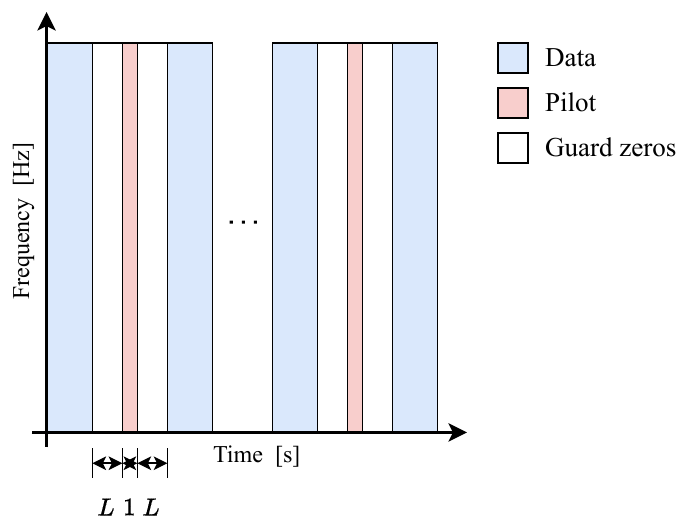}
  \caption{Separation in time, $Q+1$ pilot \\ cluster repetitions}
  \label{fig:TDKD}
\end{subfigure}%
\begin{subfigure}{.33\textwidth}
  \centering
  \includegraphics[width=1\textwidth]{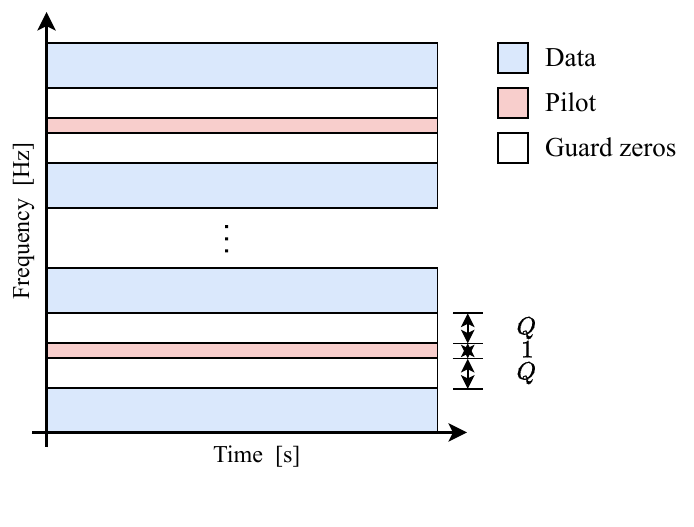}
  \caption{Separation in frequency, $L+1$ pilot \\ cluster repetitions}
  \label{fig:FDKD}
\end{subfigure}%
\caption{Two methods, as proposed by (a)\cite{ma2003optimal,kannu2005mseoptimal,kannu2008design} and (b)\cite{kannu2005mseoptimal,kannu2008design}.}
\label{fig:LTVOptimalAllocations}
\vspace{-0.5cm}
\end{figure*}

Although the studies mentioned above on LTI channels typically assume SCM or OFDM modulation, studies on LTV channels do not impose any specific assumptions on the modulation scheme employed. However, we will see that in these studies the pilot symbols and communications symbols are usually separated in either time or frequency, which links directly to SCM and OFDM, respectively.
In \cite{ma2003optimal} it was shown that, assuming that the pilot symbols and communication symbols to be transmitted are separated in time, equispaced (in time) and equipowered clusters of pilot symbols are optimal in terms of (a lower bound on) the average channel capacity. The optimal length of a cluster is $2L+1$, where the leading and trailing $L$ symbols are set to zero. In total, $Q+1$ pilot groups should be placed (equispaced) in time. Note that this matches the Nyquist sampling theorem, which tells us that the channel should be sampled twice the maximum Doppler shift (which is $Q/2$). The approach suggested in \cite{ma2003optimal} is essentially to probe the channel $Q+1$ times over a specific time period, the probe signal being an impulse. This is visualized in Fig. \ref{fig:TDKD}. Note that this approach coincides with the work on SCM for LTI channels; if we set $Q=0$ we probe the channel once with a single cluster of pilot symbols, where the leading and trailing $L$ symbols are set to zero.
Furthermore, similar to the work in \cite{adireddy2002optimal,ohno2002optimal}, the authors derived the optimal power distribution, leading to a significant improvement in the (average) channel capacity~\cite{ma2003optimal}.

Independently, \cite{kannu2005mseoptimal} proposes three pilot designs that are optimal in terms of MSE on the channel taps. The first separates the pilot and communication symbols in time and is, in fact, equivalent to the one proposed in \cite{ma2003optimal}. The second design separates the pilot and the communication symbols in frequency. 
Interestingly, this second scheme also uses guard symbols (zeros); more specifically, it inserts just enough guard symbols so that the pilot and data parts do not overlap in frequency after passing through the channel. This design probes the channel $L+1$ times in frequency, using a probe signal that consists of a single frequency surrounded by $Q$ zero frequencies on either side. The resulting pilot design has pilot symbols that are equispaced in frequency and equipowered. The pilot design is visualized in Fig. \ref{fig:FDKD}. Note again that this approach subsumes the work on OFDM for LTI channels; if we set $Q=0$, the channel is probed with $L+1$ frequencies.
The last scheme proposed by \cite{kannu2005mseoptimal,kannu2008design} is based on linear chirps; however, this scheme turned out to be less efficient in terms of bits/s/Hz \cite{kannu2008design}. Therefore, in what follows, we will not consider this pilot design.

Furthermore, it was shown that for $L>Q$, the separation in frequency achieves higher capacity, while in the case of $L<Q$, the separation in time achieves higher capacity \cite{kannu2005mseoptimal,kannu2008design}.

On a final note, for the allocations in Fig. \ref{fig:LTVOptimalAllocations}, in terms of MSE, in case of separation in time/frequency, it does not matter where the pilot blocks are in time/frequency as long as they are equispaced.

The authors of \cite{li2012modified} analyzed the case where Doppler spread is less than expected/modeled, i.e. the last few channel taps for high Doppler shifts are zero. In this case, one can design different (sub-optimal) pilot allocations. By changing zeros to non-zeros in the pilot part the channel estimation can be improved. However, note that this is a sub-optimal solution; ideally one would have to change the number of symbols in the pilot and data parts according to the channel.  Similar findings were also reported in \cite{islam2011on}.

\subsection{Pilot allocation for OTFS}
\noindent In this section, we discuss the most relevant pilot allocation schemes that were proposed for OTFS. We focus exclusively on pilot allocations where there is no overlap between pilot and communication symbols, since this will be the underlying assumption that will become evident further in the paper. For information on superimposed pilot designs, readers can consult \cite{mishra2022OTFS} and references therein.

\begin{figure*}[htbp]
\begin{subfigure}{.33\textwidth}
  \centering
   \includegraphics[width=1\textwidth]{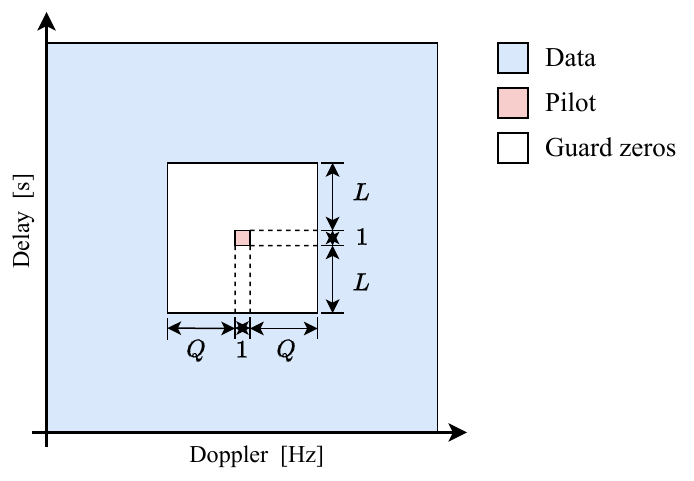}
  \caption{Integer Doppler case \cite{raviteja2018embedded,raviteja2019embedded}}
  \label{fig:integerDopplerRaviteja}
\end{subfigure}%
\begin{subfigure}{.33\textwidth}
  \centering
  \includegraphics[width=1\textwidth]{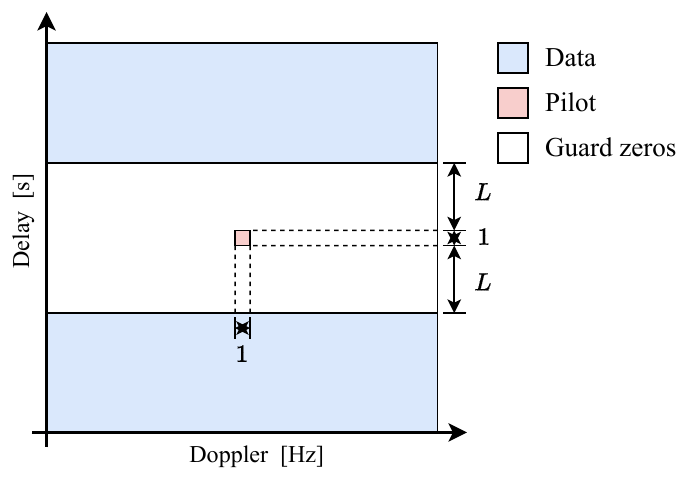}
  \caption{Fractional Doppler case \cite{raviteja2018embedded,raviteja2019embedded}}
  \label{fig:fractionalDopplerRaviteja}
\end{subfigure}%
\begin{subfigure}{.33\textwidth}
  \centering
    \includegraphics[width=1\textwidth]{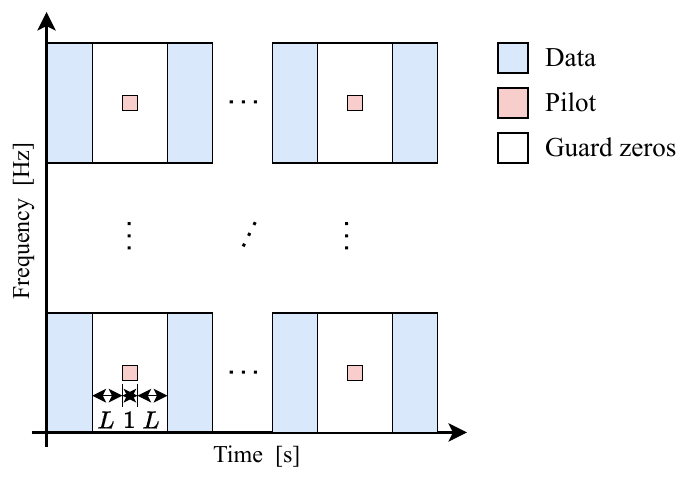}
  \caption{Time-frequency plot of (b).}
  \label{fig:TFfractional}
\end{subfigure}
\caption{Pilot allocations in the delay-Doppler and time-frequency domain.}
\label{fig:DDAllocations}
\vspace{-0.5cm}
\end{figure*}

In \cite{raviteja2018embedded,raviteja2019embedded} two pilot allocations are proposed. The first places one pilot surrounded by zeros in the delay-Doppler (DD) domain, so that no interference is possible between the pilot and the communication symbols at the receiver side. The second allocation is based on the observation that, in many real-life scenarios, the actual Doppler shifts do not perfectly match the Doppler grid of the modulation, causing fractional Doppler shifts. Therefore, the second allocation scheme also proposes to use one pilot surrounded by zeros in the DD domain, but now all Doppler bins next to the pilot are set to zero. The two schemes are visualized in Fig. \ref{fig:integerDopplerRaviteja} and Fig. \ref{fig:fractionalDopplerRaviteja}. We make some remarks here.

First of all, note that no optimality analysis is performed. The work in \cite{raviteja2019OTFS} briefly describes \textit{empirical} results of BER versus SNR of the pilot, but does not consider the power balance between pilot and communication symbols. 

Secondly, in our previous research \cite{werf2024onthe} it has been established that the symbols present in the DD domain will exhibit repetition in the time-frequency domain. Note that the repetitions in time were also depicted in Fig. \ref{fig:OTFSmodvis}. The repetitions in both time and frequency are illustrated in Fig. \ref{fig:TFfractional}, where the time-frequency plot of the signal in Fig. \ref{fig:fractionalDopplerRaviteja} is shown. This raises the question of how many repetitions would be optimal.

Thirdly, the time-frequency plot in Fig. \ref{fig:TFfractional} shows great similarity to the pilot design in Fig. \ref{fig:TDKD}. Let $\mathbf{S}_\textrm{DD}$ be the symbols in the delay-Doppler domain, then the transmitted signal is given by $\mathbf{x} = \textrm{vec}(\mathbf{S}_\textrm{DD}\mathbf{F}^H)$~\cite{raviteja2019OTFS,werf2024onthe}. This is visualized in Fig. \ref{fig:OTFSmodvis} for $N = 3$ and $M = 6$. It is clear that the transmitted signal in time contains $N$ clusters of pilot symbols, where in each cluster the leading and trailing $L$ symbols are zero. We can conclude that this overlaps with the method in \cite{ma2003optimal} and \cite{kannu2005mseoptimal,kannu2008design}. Both methods separate the pilot and the data in time and use the same size of clusters of pilot symbols (i.e. $2L+1$). If $N=Q+1$, even the number of repetitions of the clusters of pilot symbols is equivalent. The difference from the design proposed by \cite{ma2003optimal,kannu2005mseoptimal,kannu2008design} is that the impulses have different phases. The result is that for the OTFS pilot scheme, the pilots are not active on all frequencies, while in \cite{ma2003optimal,kannu2005mseoptimal,kannu2008design} it is unclear what the frequency behavior of the pilot clusters is. 

It is worth mentioning that \cite{ma2003optimal} did make an interesting remark that by having repetitions in time, one must also have some repetition in frequency\footnote{Quote: \textit{``We wish to show in this subsection that our optimal PSAM (...) enables 2-D sampling and estimation of our time-frequency selective channel. Intuitively thinking, the Kronecker deltas (...) surrounded by zero-guards implement time-domain sampling with pilot symbols; furthermore, the fact that these deltas are periodically inserted implies that they are also equivalent to Kronecker deltas in the frequency-domain and thus serve as pilot tones as well."}}. Secondly, \cite{ma2003optimal,kannu2005mseoptimal,kannu2008design} do not specify how the communication symbols are distributed, while \cite{raviteja2018embedded,raviteja2019embedded} uses the OTFS modulation to ``precode" the communication symbols. In fact, \cite{ma2003optimal} does not consider the structure and coding scheme of the communication symbols.

A peak-to-average power ratio (PAPR) analysis for OTFS is performed in \cite{marsalek2019orthogonal}. It was shown that the PAPR increases significantly when the pilot power is increased, unlike the unfounded statement made in \cite{raviteja2018embedded} saying that due to the ``spread-spectrum nature of OTFS", one can increase the pilot power without increasing the PAPR. 
Unfortunately, the details of the pilot design in  \cite{marsalek2019orthogonal} (for example, the exact size of the guard symbols) are missing.
Although the optimal point (lowest BER) of the power distribution is found \textit{empirically}, no mathematical analysis is given, making it difficult to generalize the results to different scenarios. 

\begin{figure*}
\centering
\begin{minipage}{.66\textwidth}
\centering
\begin{subfigure}{.5\textwidth}
  \centering
   \includegraphics[width=1\textwidth]{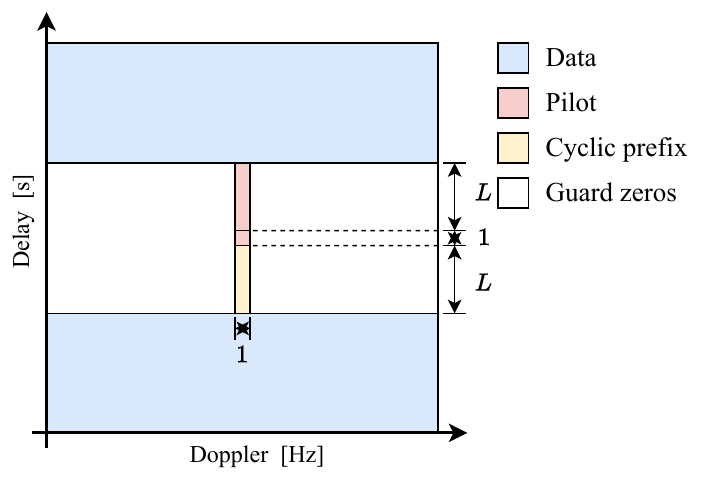}
  \caption{Delay-Doppler domain}
  \label{fig:DDcppilotdelay}
\end{subfigure}%
\begin{subfigure}{.5\textwidth}
  \centering
  \includegraphics[width=1\textwidth]{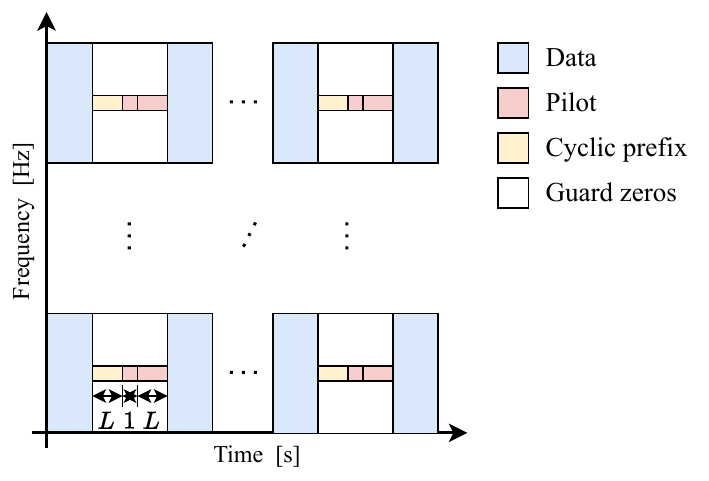}
  \caption{Time-frequency domain}
  \label{fig:TFcppilotdelay}
\end{subfigure}
\caption{Pilot allocation as proposed by \cite{sanoopkumar2023apractical}.}
\label{fig:CPPilotDelay}
\end{minipage}%
\begin{minipage}{0.33\textwidth}
  \centering
  \includegraphics[width=\linewidth]{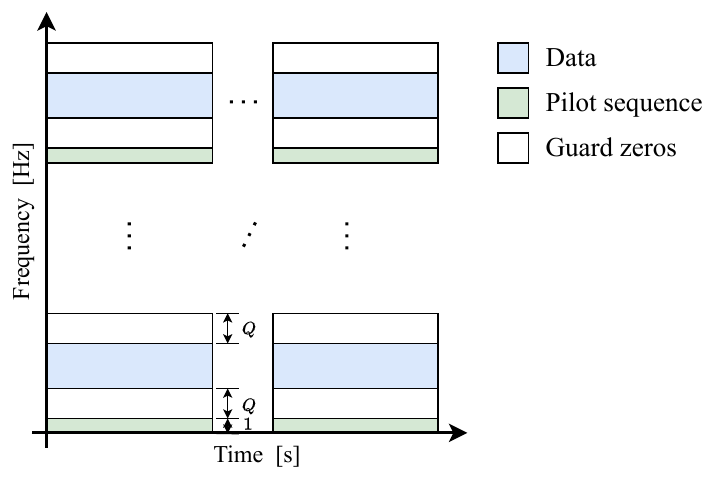}
  %\captionof{figure}{Another figure}
  \captionof{figure}{Time-frequency plot for D-OSDM \cite{ebihara2016dopplerresilient}.}
  \label{fig:DOSDM}
\end{minipage}
\vspace{-0.5cm}
\end{figure*}

Another work that proposed to reduce the PAPR of OTFS/OSDM with a so-called impulse pilot was \cite{sanoopkumar2023apractical}. In this work, the authors propose to alter the allocation proposed by \cite{raviteja2018embedded} by extending the pilot along the delay axis (a Zadoff-Chu sequence is used) and inserting a CP. The resulting signal in the delay-Doppler and time-frequency domains is visualized in Fig. \ref{fig:CPPilotDelay}. The PAPR is significantly reduced, however, at the cost of a loss in BER~\cite{sanoopkumar2023apractical}.

In \cite{qu2021lowdimensional} a pilot allocation scheme is proposed where multiple pilots are placed at the four corners of the DD domain, with a certain number of guard symbols. However, it is not clear why this allocation is chosen over the allocation proposed by \cite{raviteja2018embedded}. A comparison with \cite{raviteja2018embedded} is included, however, due to different channel estimators, the results can not be compared in a fair way.

The fractional Doppler pilot allocation scheme of \cite{raviteja2018embedded} has a rather large pilot overhead. To combat this, a different method was proposed by \cite{liu2022nearoptimal}. Instead of the delay-Doppler channel model, \cite{liu2022nearoptimal} uses the GCE-BEM. Furthermore, a two-step procedure is proposed where in the second step the GCE-BEM order $Q$ is increased and $R = 1$ is set to $R =2$. In the second step the BEM coefficients and communication symbols are iteratively recomputed, also taking into account the (first guess of the) communication symbols that were demodulated in the first step. By increasing the model order, the modeling error is reduced in the second step, hence the channel estimation performance is increased. Moreover, since in the first step a smaller BEM order $Q$ is chosen, the pilot overhead decreases compared to the one-step approach of \cite{raviteja2018embedded}.

To summarize the work on pilot design for OTFS; a theoretical analysis on optimal pilot design (allocation and power distribution) is still lacking, while the work on LTV channels suggest that such an optimization can significantly improve the performance of the modulation.

\subsection{Pilot allocation for OSDM}
\noindent In \cite{ebihara2014underwater}, reasoning from a ``sequence point of view", the authors propose to devote one sequence of length $M$ out of the $N$ sequences to pilot symbols. They use a sequence (of symbols) that is shift orthogonal. Guard intervals between the pilot and the data symbols were not used. Later, in \cite{ebihara2016dopplerresilient} this idea was generalized to a modulation scheme called D-OSDM, including guard intervals; see visualization in Fig. \ref{fig:DOSDM}.
Note that while both \cite{ebihara2016dopplerresilient} and \cite{sanoopkumar2023apractical} proposed to use a shift orthogonal sequence, \cite{ebihara2016dopplerresilient} separates the pilot and communication symbols in frequency while \cite{sanoopkumar2023apractical} separates it in time. This similarity suggests that D-OSDM will perform better in terms of PAPR than the impulse pilot OTFS/OSDM~\cite{raviteja2019embedded}, but with a compromise on BER.

Again, no analysis of optimal pilot design (allocation and power distribution) was performed. 

\section{OTFS/OSDM modulation through the LTV channel}\label{sec:OTFSOSDMproperties}
\noindent In this section, we explore the interaction of the OTFS modulation scheme with a CE-BEM channel with $L$ temporal delays and $Q$ Doppler shifts. It will be observed that the modulation scheme transforms the time-varying channel into a time-invariant channel in the DD domain. Furthermore, the received signal, in the DD domain, can be represented as a circularly shifted form of the transmitted signal (in the same domain). 

If we receive $K$ samples over time, where $K$ is factored as $K = NM$, the received signal can be expressed as,% given in \eqref{eq:rOTFSrewritten}.
% at the bottom of page 6.
% %
% \begin{figure*}[b]
% \hrulefill
\begin{equation}
\begin{split}
    \mathbf{r} 
    &
    = \mathbf{H}(\mathbf{F}^H_N \otimes \mathbf{I}_M)\mathbf{s} +\mathbf{n}
    \\
    & = \left( \sum_{q=-Q/2}^{Q/2} \sum_{l=0}^L c_{l,q} \bm{\Lambda}^{(q)}_K \mathbf{P}_K^l \right)(\mathbf{F}^H_N \otimes \mathbf{I}_M)\mathbf{s} +\mathbf{n} \\
    & = \sum_{q=-Q/2}^{Q/2} \sum_{l=0}^L c_{l,q} (\bm{\Lambda}^{(q)}_N \otimes \bm{\Lambda}^{(q/N)}_M) \overbrace{\left[(\mathbf{I}_N \otimes \mathbf{L}_M^{(l)})+(\mathbf{P}_N \otimes \mathbf{U}_M^{(l)})\right]}^{\mathbf{P}^l_K} (\mathbf{F}^H_N \otimes \mathbf{I}_M)\mathbf{s} +\mathbf{n}\\
    & = \sum_{q=-Q/2}^{Q/2} \sum_{l=0}^L c_{l,q} \left[(\bm{\Lambda}^{(q)}_N \mathbf{I}_N \otimes \bm{\Lambda}^{(q/N)}_M \mathbf{L}_M^{(l)}) +(\bm{\Lambda}^{(q)}_N \mathbf{P}_N \otimes \bm{\Lambda}^{(q/N)}_M\mathbf{U}_M^{(l)})\right](\mathbf{F}^H_N \otimes \mathbf{I}_M)\mathbf{s} +\mathbf{n}\\
    & = \sum_{q=-Q/2}^{Q/2} \sum_{l=0}^L c_{l,q} \left[(\bm{\Lambda}^{(q)}_N \mathbf{F}^H_N \otimes \bm{\Lambda}^{(q/N)}_M \mathbf{L}_M^{(l)}) +(\bm{\Lambda}^{(q)}_N \mathbf{P}_N \mathbf{F}^H_N \otimes \bm{\Lambda}^{(q/N)}_M\mathbf{U}_M^{(l)})\right]\mathbf{s} +\mathbf{n}.\\
\end{split}    
\label{eq:rOTFSrewritten}
\end{equation}
% \vspace{-0.5cm}
% \end{figure*}
%
At the receiver side we first demodulate the signal, 
% \eqref{eq:yOTFSrewritten} at the bottom of page 6.
%
% \begin{figure*}[b]
% \hrulefill
\begin{equation}
\begin{split}
    \mathbf{y} 
    & = (\mathbf{F}_N \otimes \mathbf{I}_M)\mathbf{r} 
    \\ &
    = \sum_{q=-Q/2}^{Q/2} \sum_{l=0}^L c_{l,q} \left[(\mathbf{F}_N \bm{\Lambda}^{(q)}_N  \mathbf{F}^H_N \otimes \bm{\Lambda}^{(q/N)}_M \mathbf{L}_M^{(l)}) +(\mathbf{F}_N \bm{\Lambda}^{(q)}_N \mathbf{P}_N \mathbf{F}^H_N \otimes \bm{\Lambda}^{(q/N)}_M\mathbf{U}_M^{(l)})\right]\mathbf{s} +(\mathbf{F}_N \otimes \mathbf{I}_M)\mathbf{n}.\\
\end{split}
\label{eq:yOTFSrewritten}
\end{equation}
% \vspace{-0.5cm}
% \end{figure*}
%
Then, if the vector $\mathbf{y}$ is reshaped into an $M\times N$ matrix we can write,% \eqref{eq:DDresponse},
%
% \begin{figure*}[b]
% \hrulefill
\begin{align}
% \begin{split}
    \mathbf{Y} 
    &
    = \textrm{vec}^{-1}(\mathbf{y}) 
    \notag\\ 
    & = \sum_{q=-Q/2}^{Q/2} \sum_{l=0}^L c_{l,q} \bm{\Lambda}^{(q/N)}_M \left[\mathbf{L}^{(l)}_M \mathbf{S}\mathbf{F}_N^H \bm{\Lambda}^{(q)}_N \mathbf{F}_N + \mathbf{U}^{(l)}_M \mathbf{S}\mathbf{F}_N^H \mathbf{P}_N \bm{\Lambda}^{(q)}_N \mathbf{F}_N\right] + \mathbf{N}\mathbf{F}_N \notag\\
    & = \sum_{q=-Q/2}^{Q/2} \sum_{l=0}^L c_{l,q} \bm{\Lambda}^{(q/N)}_M \left[\mathbf{L}^{(l)}_M \mathbf{S}\mathbf{P}^{-q}_N + \mathbf{U}^{(l)}_M \mathbf{S}\mathbf{F}_N^H \mathbf{P}_N \bm{\Lambda}^{(q)}_N \mathbf{F}_N\right] + \mathbf{N}\mathbf{F}_N \notag\\
    & = \sum_{q=-Q/2}^{Q/2} \sum_{l=0}^L c_{l,q} \bm{\Lambda}^{(q/N)}_M \left[\mathbf{L}^{(l)}_M \mathbf{S}\mathbf{P}^{-q}_N + \mathbf{U}^{(l)}_M \mathbf{S}\mathbf{P}^{-q}_N\textrm{diag}\left(e^{-j2\pi (0-q)/N},\dots,e^{-j2\pi ((N-1)-q)/N}\right)\right] + \mathbf{N}\mathbf{F}_N \notag\\
    %& = \textrm{\textcolor{red}{...}} \\
    & = \sum_{q=-Q/2}^{Q/2} \sum_{l=0}^L c_{l,q} \mathbf{W}_{l,q} \circ \left(\mathbf{P}_M^l \mathbf{S}\mathbf{P}_N^{-q}\right) + \mathbf{N}\mathbf{F}_N,\label{eq:DDresponse}
% \end{split}
\end{align}
% \vspace{-0.5cm}
% \end{figure*}
%
where
\begin{equation}
    [\mathbf{W}_{l,q}]_{m,n} = 
    \begin{cases}
        e^{j2\pi q m/K}, \quad\quad\quad\quad\quad\quad \textrm{if } m\geq l, \\
        e^{j2\pi q m/K}e^{-j2\pi (n-q)/N}, \:\: \textrm{if } m<l,
    \end{cases}
\end{equation}
for $m = 0,1,\dots,M-1$, $n = 0,1,\dots,N-1$.
From \eqref{eq:DDresponse}, we learn that the delay tap $l$ (circularly) shifts the rows of $\mathbf{S}$. Similarly, the Doppler tap $q$ (circularly) shifts the columns in $\mathbf{S}$. Therefore, the first and second dimensions of $\mathbf{S}$ are often called the delay and Doppler dimensions, respectively. The channel coefficient $c_{l,q}$ and the matrix $\mathbf{W}_{l,q}$ apply an attenuation in amplitude and phase.
In total, the LTV channel is ``scrambling" the transmitted symbols.
%(NB: $\mathbf{S} = \textrm{vec}^{-1}(\mathbf{s}_\textrm{c}+\mathbf{s}_\textrm{p})$) with respect to the rows and columns, respectively.

\section{Pilot distribution for OTFS}
\noindent In this section, we will examine the pilot design for OTFS modulation, taking into account the relationship between the channel and the transmitted signal, as derived in the previous section. The following steps will be followed:
\begin{enumerate} 
    \item First, we assume no overlap exists between the pilot and communication symbols at the receiver side (and thus also at the transmitter side) and analyze the pilot allocations satisfying this assumption. (Section \ref{subsec:Step1})
    \item Then we derive the allocations with minimum number of pilot overhead. (Section \ref{subsec:lowestoverhead})
    \item We show that for fixed pilot power, these allocations achieve a channel estimate that minimizes the mean squared error (MSE) on channel taps. (Section \ref{subsec:MMSEoptimality})
    \item Finally, we optimize the power distribution between the pilot and the communication symbols with respect to a performance measure. (Section \ref{subsec:optimizingPowerBalance})
\end{enumerate}

\subsection{Step 1) Analyzing possible pilot allocations}
\label{subsec:Step1}
\noindent We will assign the available symbols to pilot and communication symbols. Let $K_\textrm{c}$ and $K_\textrm{p}$ denote the number of pilot and communication symbols, such that $K = K_\textrm{c}+K_\textrm{p}$. In matrix vector notation, this division can be written as $\mathbf{s} = (\bm{\Phi}_\textrm{c}\mathbf{s}_\textrm{c} + \bm{\Phi}_\textrm{p}\mathbf{s}_\textrm{p})$. Here, $\bm{\Phi}_\textrm{c} \in\{0,1\}^{K\times K_\textrm{c}}$ and $\bm{\Phi}_\textrm{p} \in\{0,1\}^{K\times K_\textrm{p}}$ are selection matrices containing only $K_\textrm{c}$ and $K_\textrm{p}$ columns of $\mathbf{I}_K$, indexed by $\mathbf{p}_\textrm{c}\in\{0,1\}^{K\times 1}$ and $\mathbf{p}_\textrm{p}\in\{0,1\}^{K\times 1}$. We have $\mathbf{1}_{K\times 1}^T\mathbf{p}_\textrm{c} = K_\textrm{c}$ and $\mathbf{1}_{K\times 1}^T\mathbf{p}_\textrm{p} = K_\textrm{p}$, and a symbol is either used for the pilot or communication, hence $\mathbf{p}_\textrm{c} + \mathbf{p}_\textrm{p} = \mathbf{1}_{K\times 1}$.

After demodulation we can write our received signal as,
\begin{equation*}
\begin{split}
    \mathbf{y}
    & = (\mathbf{F}_N \otimes \mathbf{I}_M)\mathbf{H}(\mathbf{F}^H_N \otimes \mathbf{I}_M)(\bm{\Phi}_\textrm{c}\mathbf{s}_\textrm{c} + \bm{\Phi}_\textrm{p}\mathbf{s}_\textrm{p}) 
    % \\
    % & \hspace{5cm}
    + (\mathbf{F}_N \otimes \mathbf{I}_M)\mathbf{n}
    \\
    & 
    = \mathbf{H}_\textrm{DD}(\bm{\Phi}_\textrm{c}\mathbf{s}_\textrm{c} + \bm{\Phi}_\textrm{p}\mathbf{s}_\textrm{p}) +\mathbf{w}.\\
\end{split}
\end{equation*}
Here $\mathbf{H}_\textrm{DD} = (\mathbf{F}_N \otimes \mathbf{I}_M)\mathbf{H}(\mathbf{F}^H_N \otimes \mathbf{I}_M)$ and $\mathbf{w} = (\mathbf{F}_N \otimes \mathbf{I}_M)\mathbf{n}$.
At the receiver, we can divide the received signal into a communication part and a pilot part, where we denote the communication part by $\mathbf{y}_\textrm{c} = \bm{\Psi}^H_\textrm{c}\mathbf{y}$ and the pilot part by $\mathbf{y}_\textrm{p} = \bm{\Psi}^H_\textrm{p}\mathbf{y}$. Here, the matrices $\bm{\Psi}_\textrm{c} \in \mathbb{C}^{K\times R_\textrm{c}}$ and $\bm{\Psi}_\textrm{p}\in \mathbb{C}^{K\times R_\textrm{p}}$ denote selection matrices, similar to the definition of $\bm{\Phi}_\textrm{c}$ and $\bm{\Phi}_\textrm{p}$, and contain $R_\textrm{c}$ and $R_\textrm{p}$ columns of $\mathbf{I}_K$ indexed by $\Tilde{\mathbf{p}}_\textrm{c}\in\{0,1\}^{K\times 1}$ and $\Tilde{\mathbf{p}}_\textrm{p}\in\{0,1\}^{K\times 1}$, respectively. It is important to mention that the $\mathbf{\Psi}$ matrices are selecting all the received symbols that include a communication symbol (for $\mathbf{\Psi}_\textrm{c}$) or a pilot symbol (for $\mathbf{\Psi}_\textrm{p}$). Thus, if $\mathbf{\Phi}$ is designed, $\mathbf{\Psi}$ follows automatically. Besides, note that the selection of $\bm{\Psi}_\textrm{p}$ is larger than that of $\bm{\Phi}_\textrm{p}$, i.e. $R_\textrm{c}\geq K_\textrm{c}$, because the transmitted communication symbols are spread out by the channel.  
%In the special case where $L=0$ and $Q = 0$, $R_\textrm{c} = K_\textrm{c}$, and the matrices $\bm{\Phi}_\textrm{c}$ and $\bm{\Psi}_\textrm{c}$ are equal.
We make the following assumption.
\begin{itemize}
    \item [\textbf{A1}] There is no overlap between pilot and communication symbols at the receiver side.
\end{itemize}
It follows from \textbf{A1} that we should have $\bm{\Psi}^H_\textrm{c}\mathbf{H}_\textrm{DD}\bm{\Phi}_\textrm{p} = \mathbf{0}$ and $\bm{\Psi}^H_\textrm{p}\mathbf{H}_\textrm{DD}\bm{\Phi}_\textrm{c} = \mathbf{0}$. From the interaction between the channel and the modulation, as derived in \eqref{eq:DDresponse}, we know that this is possible as long as the pilot and communication symbols are guarded by zeros accordingly, so that the shift operations of the channel do not mix them. 

The communication part is given by
\begin{equation}
\begin{split}
    \mathbf{y}_\textrm{c} 
    = \bm{\Psi}^H_\textrm{c}\mathbf{y} & = \bm{\Psi}^H_\textrm{c}\mathbf{H}_\textrm{DD}(\bm{\Phi}_\textrm{c}\mathbf{s}_\textrm{c} + \bm{\Phi}_\textrm{p}\mathbf{s}_\textrm{p}) +\bm{\Psi}^H_\textrm{c}\mathbf{w} \\
    & = \bm{\Psi}^H_\textrm{c}\mathbf{H}_\textrm{DD}\bm{\Phi}_\textrm{c}\mathbf{s}_\textrm{c}  +\mathbf{w}_\textrm{c},
\end{split}
\end{equation}
and that the pilot part is given by
\begin{equation}
\begin{split}
    \mathbf{y}_\textrm{p} 
    =  \bm{\Psi}^H_\textrm{p}\mathbf{y} & = \bm{\Psi}^H_\textrm{p}\mathbf{H}_\textrm{DD}(\bm{\Phi}_\textrm{c}\mathbf{s}_\textrm{c} + \bm{\Phi}_\textrm{p}\mathbf{s}_\textrm{p}) +\bm{\Psi}^H_\textrm{p}\mathbf{w} \\
    & = \bm{\Psi}^H_\textrm{p}\mathbf{H}_\textrm{DD}\bm{\Phi}_\textrm{p}\mathbf{s}_\textrm{p}  +\mathbf{w}_\textrm{p}.\\
\end{split}
\label{eq:yp}
\end{equation}

\begin{figure*}[htbp]
\begin{subfigure}{.33\textwidth}
  \centering
   \includegraphics[width=1\textwidth]{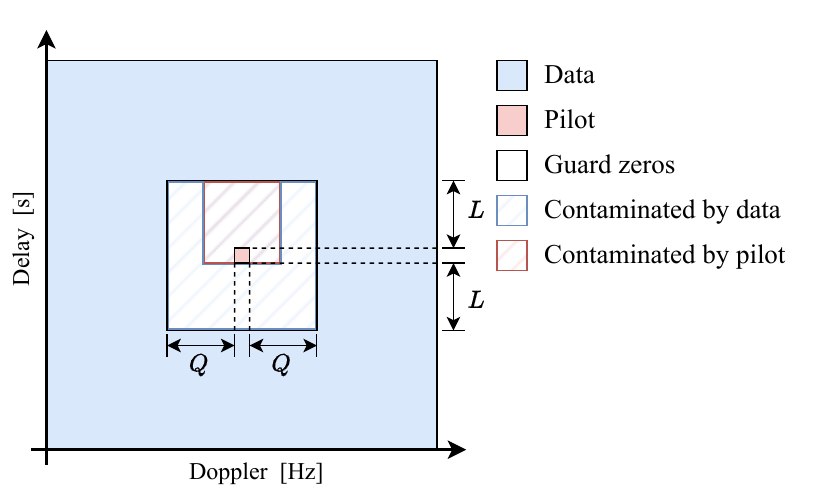}
  \caption{$M\geq2L+1$ and $N \geq 2Q+1$}
  \label{fig:Case1}
\end{subfigure}%
\begin{subfigure}{.33\textwidth}
  \centering
  \includegraphics[width=1\textwidth]{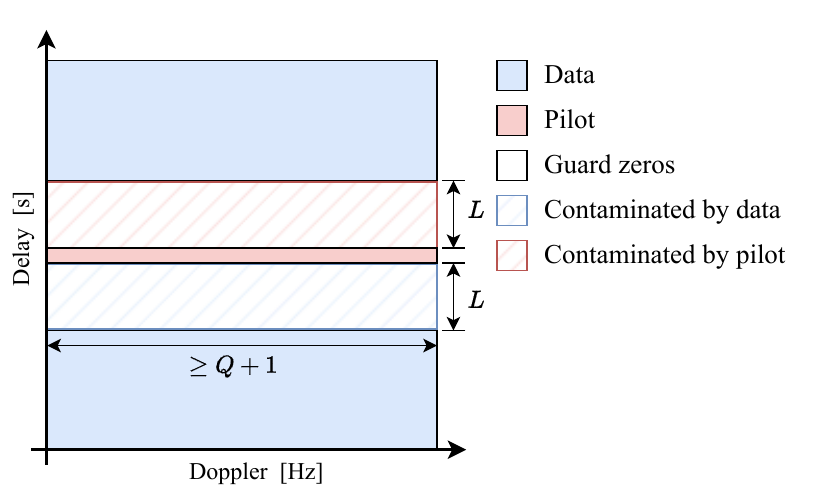}
  \caption{$M\geq 2L+1$ and $N \geq Q+1$}
  \label{fig:Case2}
\end{subfigure}%
\begin{subfigure}{.33\textwidth}
  \centering
    \includegraphics[width=1\textwidth]{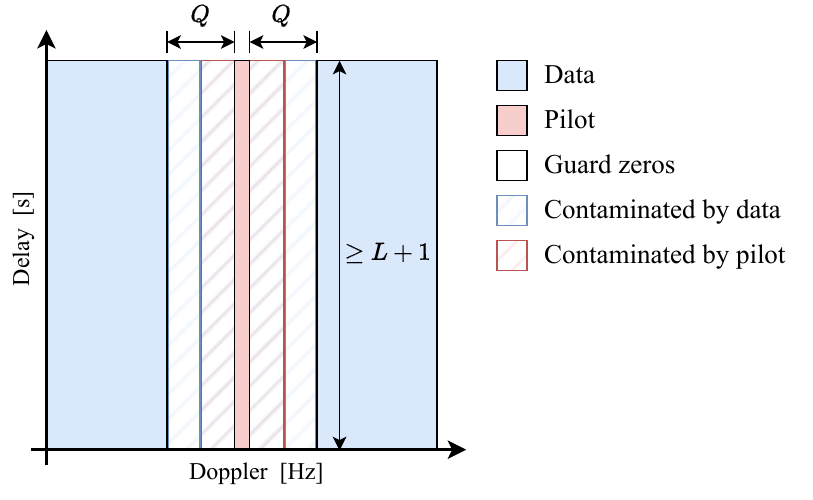}
  \caption{$M\geq L+1$ and $N \geq 2Q+1$}
  \label{fig:Case3}
\end{subfigure}
\caption{DD configurations satisfying \textbf{A1}.}
\label{fig:threecases}
\vspace{-0.5cm}
\end{figure*}

Recognizing that the LTV channel (circularly) shifts the rows and columns of the transmitted symbols $\mathbf{S}_\textrm{DD}$, we can deduce which type of pilot design respects \textbf{A1}. We identify that every pilot design must consist of one or a combination of the following three ``basic" cases.
\begin{enumerate}
    \item ``Island case" - in case the pilot symbol ``area" is embedded both in the delay ánd Doppler direction by communication symbols, in order to have nonzero pilot symbols, the ``area" should be at least $(2Q+1)\times(2L+1)$, thus $K_\textrm{p} \geq (2Q+1)(2L+1)$. See Fig. \ref{fig:Case1} for a visualization. Note that this pilot allocation requires $N\geq 2Q+1$ and $M\geq 2L+1$.
    \item ``Doppler slab" - in case the pilot symbol ``area" is embedded only in the delay direction but not in the Doppler direction, the ``area should be at least $N\times(2L+1)$, thus $K_\textrm{p} \geq N(2L+1)$. See Fig. \ref{fig:Case2} for a visualization. Note that this pilot allocation requires $N\geq Q+1$ and $M\geq 2L+1$.
    \item ``Delay slab" - in case the pilot symbol ``area" is embedded only in the Doppler direction but not in the delay direction, the ``area should be at least $(2Q+1)\times M$, thus $K_\textrm{p} \geq (2Q+1)M$. See Fig. \ref{fig:Case3} for a visualization. Note that this pilot allocation requires $N\geq 2Q+1$ and $M\geq L+1$.
\end{enumerate}

Having determined the possibilities for the pilot design, we are interested in its symbol overhead and the estimation performance. 

\subsection{Step 2) Determining the pilot allocations with the lowest overhead}\label{subsec:lowestoverhead}
\noindent The minimum number of pilot symbols for each case is given by
\begin{enumerate}
    \item ``Island case" - $K_\textrm{p} = (2Q+1)(2L+1)$, consequently only one pilot symbol is nonzero.
    \item ``Doppler slab" - $K_\textrm{p} = (Q+1)(2L+1)$, thus choosing $N = Q+1$, consequently only one row containing $Q+1$ pilot symbols is nonzero.
    \item ``Delay slab" - $K_\textrm{p} = (2Q+1)(L+1)$, thus choosing $M = L+1$, consequently only one column containing $L+1$ pilot symbols is nonzero.
\end{enumerate}
It is clear that case 2) and case 3) will require less pilot overhead compared to case 1), for every choice of $L$ and $Q$. Therefore, it is crucial to examine whether there are any performance distinctions among these pilot designs. In the subsequent section, we will demonstrate that all these designs (1), 2), and 3)) achieve the same minimum MSE.

\subsection{Step 3) Showing MMSE optimality} \label{subsec:MMSEoptimality}
\noindent The pilot part at the receiver side can be rewritten as
\begin{equation}
    \mathbf{y}_\textrm{p} = \bm{\Psi}^H_\textrm{p}\mathbf{H}_\textrm{DD}\bm{\Phi}_\textrm{p}\mathbf{s}_\textrm{p} +\mathbf{w}_\textrm{p} = \mathbf{Z}\mathbf{c} +\mathbf{w}_\textrm{p},
\end{equation}
where $\mathbf{c}\in\mathbb{C}^{(L+1)(Q+1)\times 1}$ contains the coefficients $c_{l,q}$ of the BEM and where $\mathbf{Z}\in \mathbb{C}^{R_\textrm{p} \times (L+1)(Q+1)}$ is given by,% \eqref{eq:Z}.
%
% \begin{figure*}[b]
% \hrulefill
\begin{equation}
    \mathbf{Z} =
    \left[
    \bm{\Psi}^H_\textrm{p}(\mathbf{F}_N \otimes \mathbf{I}_M)\left( \bm{\Lambda}^{(q)}_K \mathbf{P}_K^l \right)(\mathbf{F}^H_N \otimes \mathbf{I}_M)\bm{\Phi}_\textrm{p}\mathbf{s}_\textrm{p}
    \right]_{l=0,\dots,L+1,\: q=-Q/2,\dots,Q/2}.
    \label{eq:Z}
\end{equation}
% \vspace{-0.5cm}
% \end{figure*}

Let $\hat{\mathbf{c}}$ represent a channel estimator. Our objective is to minimize the mean squared error (MSE) of this estimator, that is, $\mathbb{E}[(\mathbf{c}-\hat{\mathbf{c}})^H(\mathbf{c}-\hat{\mathbf{c}})]$.
Note that we have a linear model, and consequently the MSE can be minimized by the linear minimum MSE (LMMSE) estimator. Assume the channel taps are independently distributed following a Gaussian distribution with a mean of zero and potentially varying variances, such that $\mathbb{E}[\mathbf{c}\mathbf{c}^H] = \textrm{diag}([\sigma_{c_{0,0}}^2, \dots, \sigma_{c_{L,Q}}^2])=\mathbf{R}_{\mathbf{c}}$ and let $\mathbb{E}[\mathbf{w}_\textrm{p}\mathbf{w}_\textrm{p}^H] = \mathbf{R}_{\mathbf{w}_\textrm{p}}$, then, the expression for the LMMSE estimator is 
\begin{equation}
    \hat{\mathbf{c}}  = \left(\mathbf{R}^{-1}_{\mathbf{w}_\textrm{p}} \mathbf{Z}(\mathbf{R}^{-1}_{\mathbf{c}}+\mathbf{Z}^H\mathbf{R}_{\mathbf{w}_\textrm{p}}^{-1}\mathbf{Z})^{-1}\right)^H\mathbf{y}_\textrm{p}. \\
\end{equation}
If we assume that the noise is white, i.e. $\mathbf{R}_\mathbf{n} = \sigma_{\mathbf{n}}^2\mathbf{I}_{K}$, then $\mathbf{R}_\mathbf{w} = \mathbb{E}[(\mathbf{F}^H_N \otimes \mathbf{I}_M)\mathbf{n}\mathbf{n}^H(\mathbf{F}_N \otimes \mathbf{I}_M)] = \sigma_{\mathbf{n}}^2\mathbf{I}_{K}$, and consequently,
\begin{equation}
    \mathbf{R}_{\mathbf{w}_\textrm{p}} = \mathbb{E}[\bm{\Psi}^H_\textrm{p}\mathbf{w}\mathbf{w}^H\bm{\Psi}_\textrm{p}] = \bm{\Psi}^H_\textrm{p}\mathbf{R}_\mathbf{w}\bm{\Psi}_\textrm{p} = \sigma_{\mathbf{n}}^2\mathbf{I}_{R_\textrm{p}}.
\end{equation}
As a result, we can rewrite the LMMSE estimator as
\begin{equation}
\begin{split}
    \hat{\mathbf{c}} 
    & = \left(\sigma_{\mathbf{n}}^2\mathbf{R}^{-1}_{\mathbf{c}}+\mathbf{Z}^H\mathbf{Z} \right)^{-1} \mathbf{Z}^H\left(\mathbf{Z}\mathbf{c} + \mathbf{w}_\textrm{p}\right). \\
\end{split}
\end{equation}
The MSE of this estimator is given by, %\textcolor{red}{prove/show??},
\begin{equation}
\begin{split}
    \mathbb{E}[(\mathbf{c}-\hat{\mathbf{c}})^H(\mathbf{c}-\hat{\mathbf{c}})] 
    & = \mathbb{E}\left[\textrm{tr}\left((\mathbf{c}-\hat{\mathbf{c}})(\mathbf{c}-\hat{\mathbf{c}})^H\right)\right] 
    % \\
    % & 
    = 
    \textrm{tr}\left(\left( \mathbf{R}^{-1}_{\mathbf{c}} +\frac{1}{    \sigma_{\mathbf{n}}^2    }\mathbf{Z}^H\mathbf{Z}\right)^{-1}\right). \\
\end{split}
\label{eq:channelMSE}
\end{equation}

To reach the mimimum MSE, the pilot symbols should be placed such that $\mathbf{Z}^H\mathbf{Z}$ is diagonal~(\hspace{-0.007cm}\cite{ohno2004capacity}, Lemma 1).
Note that the columns of $\mathbf{Z}$ are given by
\begin{equation}
\begin{split}
    \mathbf{z}_{l+q(L+1)} 
    & = \bm{\Psi}^H_\textrm{p}(\mathbf{F}_N \otimes \mathbf{I}_M)\left( \bm{\Lambda}^{(q)}_K \mathbf{P}_K^l \right)(\mathbf{F}^H_N \otimes \mathbf{I}_M)\bm{\Phi}_\textrm{p}\mathbf{s}_\textrm{p} \\
    & = \bm{\Psi}^H_\textrm{p}\textrm{vec}\left(
    \mathbf{W}_{l,q} \circ \left(\mathbf{P}_M^l \textrm{vec}^{-1}(\bm{\Phi}_\textrm{p}\mathbf{s}_\textrm{p})\mathbf{P}_N^{-q}\right)
    \right). \\
\end{split}
\end{equation}
Let $\mathbf{S}_\textrm{p} = \textrm{vec}^{-1}(\bm{\Phi}_\textrm{p}\mathbf{s}_\textrm{p})$. The elements of the matrix $\mathbf{Z}^H\mathbf{Z}$ can be rewritten\footnote{Where (a) uses the fact that if $\mathbf{a}_1$ and $\mathbf{a}_2$ are two column vectors and $\mathbf{X}$ is a diagonal matrix, then $\mathbf{a}_1^H\mathbf{X}\mathbf{a}_2 = \textrm{vec}(\mathbf{a}^T_2 \odot \mathbf{a}_1^H)\textrm{diag}(\mathbf{X}) = (\mathbf{a}_1^\ast\circ\mathbf{a}_2)^T\textrm{diag}(\mathbf{X})$.} as in \eqref{eq:ZHZ}.
\begin{figure*}[b]
\hrulefill
\begin{equation}
\begin{split}
    & \mathbf{z}^H_{l_1+q_1(L+1)}\mathbf{z}_{l_2+q_2(L+1)} \\
    & = 
    \textrm{vec}^H
    \left(
        \mathbf{W}_{l_1,q_1} \circ \left(\mathbf{P}_M^{l_1} \mathbf{S}_\textrm{p}\mathbf{P}_N^{-q_1}\right)
    \right)
    \bm{\Psi}_\textrm{p}\bm{\Psi}_\textrm{p}^H
    \textrm{vec}
    \left(
        \mathbf{W}_{l_2,q_2} \circ \left(\mathbf{P}_M^{l_2} \mathbf{S}_\textrm{p}\mathbf{P}_N^{-q_2}\right)
    \right)\\
    & \stackrel{(a)}{=}
    \left[
    \textrm{vec}^\ast
    \left(
        \mathbf{W}_{l_1,q_1} \circ \left(\mathbf{P}_M^{l_1} \mathbf{S}_\textrm{p}\mathbf{P}_N^{-q_1}\right)
    \right)
    \circ
        \textrm{vec}
    \left(
        \mathbf{W}_{l_2,q_2} \circ \left(\mathbf{P}_M^{l_2} \mathbf{S}_\textrm{p}\mathbf{P}_N^{-q_2}\right)
    \right)
    \right]^T \textrm{diag}\left( \bm{\Psi}_\textrm{p}\bm{\Psi}_\textrm{p}^H \right)\\
    & = 
    \textrm{vec}^T
    \left(
    \mathbf{W}^\ast_{l_1,q_1}
    \circ
    \mathbf{W}_{l_2,q_2}
    \circ
    \left(\mathbf{P}_M^{l_1} \mathbf{S}_\textrm{p}^\ast \mathbf{P}_N^{-q_1}\right)
    \circ
    \left(\mathbf{P}_M^{l_2} \mathbf{S}_\textrm{p}\mathbf{P}_N^{-q_2}\right)
    \right)
    \textrm{diag}\left( \bm{\Psi}_\textrm{p}\bm{\Psi}_\textrm{p}^H \right) \\
    \end{split}
    \label{eq:ZHZ}
\end{equation}
\end{figure*}
For the diagonal elements, i.e. $(l_1,q_1) = (l_2,q_2) \in \{0,1,\dots,L\}\times\{0,1,\dots,Q\}$, we have,
\begin{equation}
\begin{split}
    \mathbf{z}^H_{i}\mathbf{z}_{i} & = \textrm{vec}^T
    \left(
    \mathbf{W}^\ast_{l,q}
    \circ
    \mathbf{W}_{l,q}
    \circ
    \left(\mathbf{P}_M^{l}
    \left(\mathbf{S}_\textrm{p}^\ast\circ\mathbf{S}_\textrm{p} \right)
    \mathbf{P}_N^{-q}\right)
    \right) 
    % \times \\
    % &  \hspace{5cm}
    \textrm{diag}\left( \bm{\Psi}_\textrm{p}\bm{\Psi}_\textrm{p}^H \right) \\
    & = \textrm{vec}^T
    \left(
    \left(\mathbf{P}_M^{l}
    \left(\mathbf{S}_\textrm{p}^\ast\circ\mathbf{S}_\textrm{p} \right)
    \mathbf{P}_N^{-q}\right)
    \right)
    \textrm{diag}\left( \bm{\Psi}_\textrm{p}\bm{\Psi}_\textrm{p}^H \right) 
    % \\
    % & 
    = \textrm{vec}^T
    \left(
    \left(\mathbf{P}_M^{l}
    \left(\mathbf{S}_\textrm{p}^\ast\circ\mathbf{S}_\textrm{p} \right)
    \mathbf{P}_N^{-q}\right)
    \right)
    \Tilde{\mathbf{p}}_\textrm{p}  \\
    & \stackrel{(b)}{=} \left(\mathbf{s}^{*}_\textrm{p} \circ \mathbf{s}_\textrm{p}\right)^T \mathbf{1}_{K_\textrm{p}\times 1} = \mathbf{s}^H_\textrm{p}\mathbf{s}_\textrm{p} = P_\textrm{p},
\end{split}
\end{equation}
where in (b) we have used the fact that $\Tilde{\mathbf{p}}_\textrm{p}$ is selecting all the symbols that include a pilot symbol after passing through the channel.
Thus, the diagonal of $\mathbf{Z}^H\mathbf{Z}$ only contains the total power $P_\textrm{p}$ of the pilot symbols. Note that the diagonal elements do not depend on $K_\textrm{p}$.
If $\mathbf{Z}^H\mathbf{Z}$ must be diagonal, then the off-diagonal elements of $\mathbf{Z}^H\mathbf{Z}$ must be zero, that is,
\begin{equation}
    \mathbf{z}^H_{l_1+q_1(L+1)}\mathbf{z}_{l_2+q_2(L+1)} = 0, \quad (l_1,q_1) \neq (l_2,q_2).
\end{equation}
By inspecting \eqref{eq:ZHZ} for $(l_1,q_1) \neq (l_2,q_2)$, we see that, left of the element-wise product, the pilot symbols are shifted by $l_1$ in the delay direction and $q_1$ in the Doppler direction, while on the right of the element-wise product, the pilot symbols are shifted by $l_2$ in the delay direction and $q_2$ in the Doppler direction. From this observation we can deduce that, to have the outcome equal to zero, \textit{the pilot symbols in the delay-Doppler domain, that is the matrix} $\textrm{vec}^{-1}(\bm{\Phi}_\textrm{p}\mathbf{s}_\textrm{p})$\textit{, should have shift orthogonal rows and columns.}

\begin{figure*}[htbp]
\centering
\begin{subfigure}{.33\textwidth}
  \centering
  \includegraphics[width=1\textwidth]{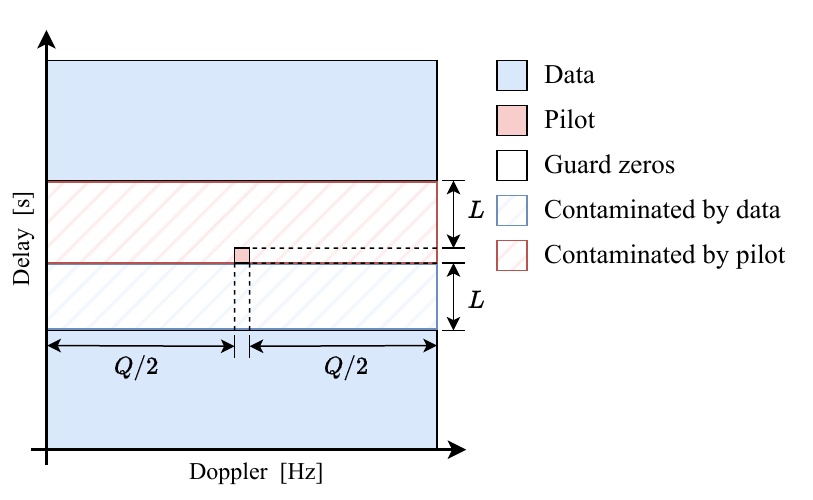}
  \caption{Doppler slab, \\ $M \geq 2L+1$, $N = Q+1$.}
  \label{fig:DopplerslabOrth}
\end{subfigure}%
\begin{subfigure}{.33\textwidth}
  \centering
    \includegraphics[width=1\textwidth]{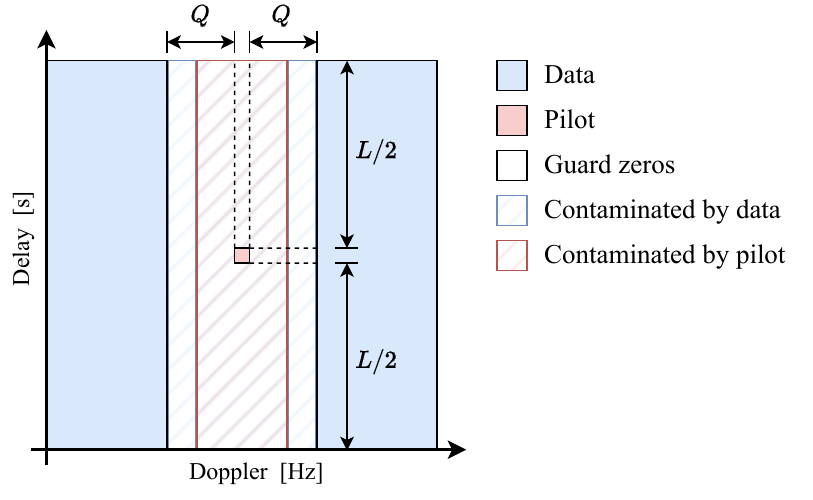}
  \caption{Delay slab, \\ $M = L+1$, $N \geq 2Q+1$.}
  \label{fig:DelayslabOrth}
\end{subfigure}
\begin{subfigure}{.33\textwidth}
  \centering
    \includegraphics[width=1\textwidth]{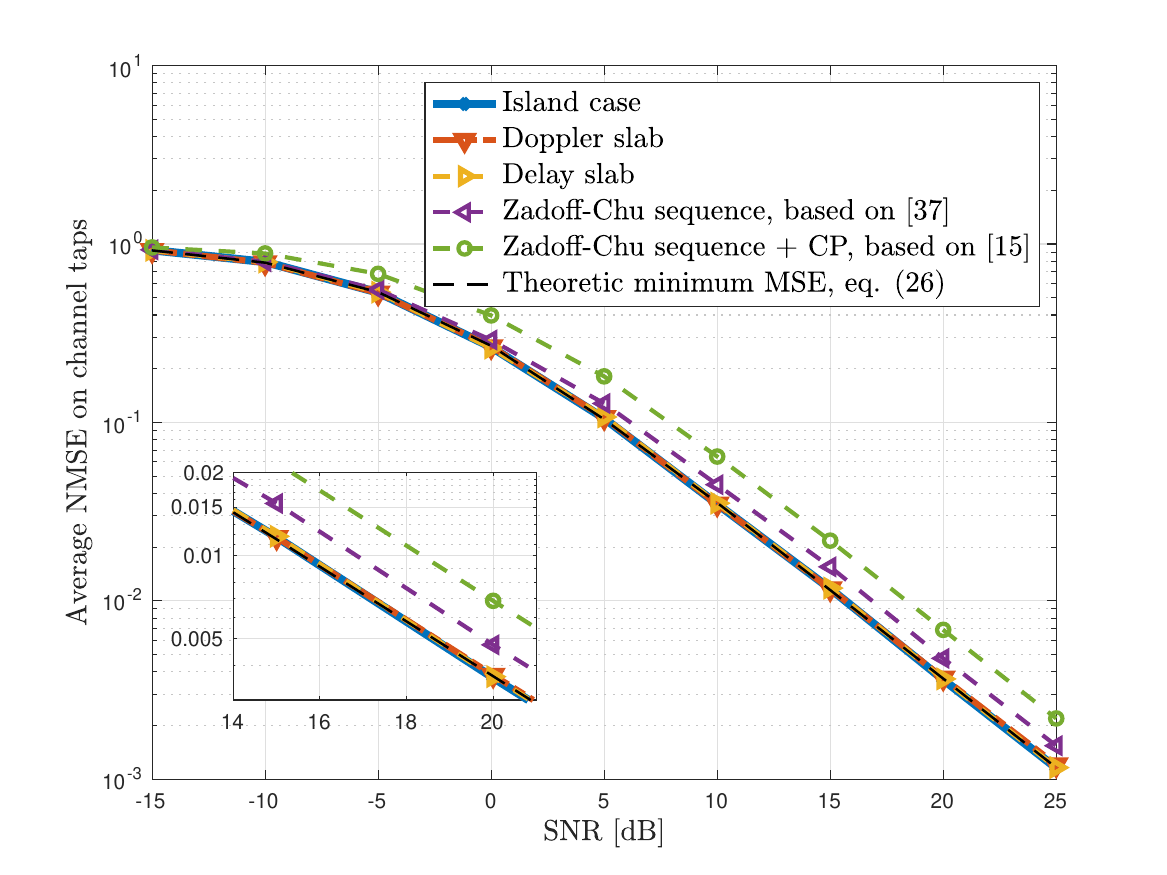}
  \caption{Channel estimation performance}
  \label{fig:ChannelEstimationPerf}
\end{subfigure}
\caption{a) and b): Pilot allocations satisfying i) \textbf{A1}, ii) achieving the MMSE ($\mathbf{Z}^H\mathbf{Z}$ diagonal), and iii) having the lowest overhead. c): Performance compared to existing pilot schemes, for $\{K,L,Q\} = \{441,8,8\}$.}
\label{fig:DDtwocasesOrth}
\vspace{-0.5cm}
\end{figure*}

We can draw some conclusions:
\begin{enumerate}
    \item ``Island case" - For a fixed power $P_\textrm{p}$, this pilot allocation, with $K_\textrm{p} = (2Q+1)(2L+1)$ and with only one nonzero pilot symbol in the middle, achieves the minimum MSE. However, it uses more pilot symbols compared to the ``Doppler slab" and ``Delay slab" case.
    \item ``Doppler slab" - For a fixed power $P_\textrm{p}$, this pilot allocation, with $K_\textrm{p} = (Q+1)(2L+1)$ and with $Q+1$ nonzero pilot symbols, achieves the minimum MSE if and only if the nonzero pilots are shift orthogonal in both delay and Doppler direction at the same time. The only option adhering to this orthogonality is to have only one nonzero pilot symbol. This is visualized in Fig. \ref{fig:DopplerslabOrth}. Although in the figure the pilot symbol is placed in the middle, the symbol could be placed anywhere along the Doppler direction (the delay position is fixed).
    \item ``Delay slab" - For a fixed power $P_\textrm{p}$, this pilot allocation, with $K_\textrm{p} = (2Q+1)(L+1)$ and with $L+1$ nonzero pilot symbols, achieves the minimum MSE if and only if the nonzero pilots are shift orthogonal in both delay and Doppler direction at the same time. The only option adhering to this orthogonality is to have only one nonzero pilot symbol. This is visualized in Fig. \ref{fig:DelayslabOrth}. Although in the figure the pilot symbol is placed in the middle, the symbol could be placed anywhere along the delay direction (the Doppler position is fixed).
\end{enumerate}
To recap, given \textbf{A1}, the pilot allocations with the lowest pilot overhead are (also) allocations that attain the minimum MSE. The latter result is corroborated by experiments, illustrated in Fig. \ref{fig:ChannelEstimationPerf}. Moreover, the three cases, the island case, Doppler slab, and delay slab, achieve the same minimum MSE; however, the Doppler slab and delay slab do so with less pilot overhead. In order to use the least overhead, in case $Q<L$ we must choose the Doppler slab and set $N = Q+1$, and in case $Q>L$ we must choose the delay slab and set $M = L+1$.

As the optimal pilot allocation is now determined, the next step is to optimize the power distribution between the pilot and the communication symbols. 

\subsection{Step 4) Optimizing the power balance}
\label{subsec:optimizingPowerBalance}
\noindent Suppose we have a total power budget $P$, and we want to distribute it between the pilot and the communications symbols. Let $P_\textrm{c} = \alpha P$ and $P_\textrm{p} =(1-\alpha)P$, where $\alpha\in[0,1]$ determines how the power is distributed, such that $P = P_\textrm{c} + P_\textrm{p}$. We can then optimize a performance measure with respect to this power balance.
Different performance measures can be chosen. Bit-error rate (BER) could be an option, however, the BER is dependent on the modulation order (i.e. BPSK, 4-QAM, 16-QAM, etc.) and on the channel coding scheme, which makes it hard to compare two configurations fairly.  
Therefore, in this work, we choose the \textit{channel capacity} as a performance measure, which does give the overall ``performance" of the communication link without having to consider the modulation order or channel coding.

The received communication part is given by,
\begin{equation}
\begin{split}
    \mathbf{y}_\textrm{c} 
    & = \bm{\Psi}^H_\textrm{c}\mathbf{H}_\textrm{DD}\bm{\Phi}_\textrm{c}\mathbf{s}_\textrm{c} +\mathbf{w}_\textrm{c} 
    = \Tilde{\mathbf{H}}_\textrm{c} \mathbf{s}_\textrm{c} + \mathbf{w}_\textrm{c}. 
\end{split}
\end{equation}
The capacity of the channel, averaged over the random channel $\Tilde{\mathbf{H}}_\textrm{c}$, induced by the coefficients in $\mathbf{c}$, is given by (see \cite{ma2003optimal,telatar1999capacity}),
\begin{equation}
    C = \frac{1}{K}\mathbb{E}\left[\max_{p(\mathbf{s}_\textrm{c}), P_\textrm{c} = \mathbb{E}[\mathbf{s}_\textrm{c}^H\mathbf{s}_\textrm{c}]}\mathcal{I}(\mathbf{y}_\textrm{c};\mathbf{s}_\textrm{c}|\hat{\mathbf{c}})\right].
\end{equation}
Here, $\mathcal{I}(\mathbf{y}_\textrm{c};\mathbf{s}_\textrm{c}|\hat{\mathbf{c}})$ is the conditional mutual information between the received signal $\mathbf{y}_\mathbf{c}$ and the transmitted symbols $\mathbf{s}_\mathbf{c}$, given the channel coefficient estimate $\hat{\mathbf{c}}$, $p(\mathbf{s}_\textrm{c})$ is the probability distribution of $\mathbf{s}_\textrm{c}$ with fixed energy $P_\textrm{c}$.

Let the channel estimate be given by $\hat{\mathbf{c}}$, and let the \textit{estimated} channel matrix be denoted by $\hat{\Tilde{\mathbf{H}}}_\textrm{c}$, then the received communication part can be rewritten as,
\begin{equation}
\begin{split}
    \mathbf{y}_\textrm{c} 
    & = \hat{\Tilde{\mathbf{H}}}_\textrm{c}\mathbf{s}_\textrm{c} +(\Tilde{\mathbf{H}}_\textrm{c}-\hat{\Tilde{\mathbf{H}}}_\textrm{c})\mathbf{s}_\textrm{c}  +\mathbf{w}_\textrm{c} 
    = \hat{\Tilde{\mathbf{H}}}_\textrm{c}\mathbf{s}_\textrm{c} + \mathbf{v}, \\
\end{split}
\end{equation}
where $\mathbf{v} = (\Tilde{\mathbf{H}}_\textrm{c}-\hat{\Tilde{\mathbf{H}}}_\textrm{c})\mathbf{s}_\textrm{c}  +\mathbf{w}_\textrm{c}$. Now, since no knowledge is available at the transmitter about the channel, it is reasonable to assign equal energy to all communication symbols, i.e. $\mathbf{R}_{\mathbf{s}_\textrm{c}} = \frac{P_\textrm{c}}{K_\textrm{c}} \mathbf{I}_{K_\textrm{c}}$.
With a fixed (i.e. equal) communication symbol power, a lower bound on the channel capacity is given by (\cite{ma2003optimal,telatar1999capacity}),
\begin{equation}
    C \geq 
    %\underline{C} = 
    \frac{1}{K}\mathbb{E}\left[ \log \det  \left(\mathbf{I}_{R_\textrm{c}\times R_\textrm{c}}+\frac{P_\textrm{c}}{K_\textrm{c}}\mathbf{R}^{-1}_\mathbf{v}\hat{\Tilde{\mathbf{H}}}_\textrm{c}\hat{\Tilde{\mathbf{H}}}_\textrm{c}^H\right)\right].
    \label{eq:Cgeq}
\end{equation}
Here,
\begin{equation}
\begin{split}
    \mathbf{R}_\mathbf{v} 
    & = \mathbb{E}[\mathbf{v}\mathbf{v}^H] 
    % \\
    % & 
    = \frac{P_\textrm{c}}{K_\textrm{c}}\mathbb{E}[(\Tilde{\mathbf{H}}_\textrm{c}-\hat{\Tilde{\mathbf{H}}}_\textrm{c})(\Tilde{\mathbf{H}}_\textrm{c}-\hat{\Tilde{\mathbf{H}}}_\textrm{c})^H]
    +
    \sigma_{\mathbf{n}}^2 \mathbf{I}_{R_\textrm{c}\times R_\textrm{c}}.
    \end{split}
    \label{eq:Rv}
\end{equation}
The goal is to allocate the right power, thus to choose $\alpha$, such that the lower bound on $C$ is maximized.

First of all, we can write (see Appendix \ref{app:derivation} for the derivation),
\begin{equation}
    \mathbb{E}[(\Tilde{\mathbf{H}}_\textrm{c}-\hat{\Tilde{\mathbf{H}}}_\textrm{c})(\Tilde{\mathbf{H}}_\textrm{c}-\hat{\Tilde{\mathbf{H}}}_\textrm{c})^H] \preceq \mathbb{E}\left[\textrm{tr}\left((\mathbf{c}-\hat{\mathbf{c}})(\mathbf{c}-\hat{\mathbf{c}})^H\right)\right] \mathbf{I}_{ R_\textrm{c}}.
    \label{eq:approximation}
\end{equation}

Secondly, because the pilot allocation makes $\mathbf{Z}^H\mathbf{Z}$ diagonal, we can write
\begin{equation}
\begin{split}
    %& 
    \left( \mathbf{R}^{-1}_{\mathbf{c}} +\frac{1}{    \sigma_{\mathbf{n}}^2    }\mathbf{Z}^H\mathbf{Z}\right)^{-1} 
    % = 
    % \\
    & 
    = \left( \textrm{diag}([\sigma_{c_{0,0}}^2, \dots, \sigma_{c_{L,Q}}^2]) +\frac{1}{    \sigma_{\mathbf{n}}^2    }P_\textrm{p}\mathbf{I}_{(L+1)(Q+1)}\right)^{-1} \\
    & = \textrm{diag}\left(\frac{\sigma_{c_{0,0}}^2 \sigma_{\mathbf{n}}^2}{\sigma_{\mathbf{n}}^2 + \sigma_{c_{0,0}}^2 P_\textrm{p}},
    \dots, 
    \frac{\sigma_{c_{L,Q}}^2 \sigma_{\mathbf{n}}^2}{\sigma_{\mathbf{n}}^2 + \sigma_{c_{L,Q}}^2 P_\textrm{p}}\right),
\end{split}
\end{equation} 
so that, the channel MSE \eqref{eq:channelMSE} is given by,
\begin{equation}
\begin{split}
    \mathbb{E}\left[\textrm{tr}(\mathbf{c}-\hat{\mathbf{c}})(\mathbf{c}-\hat{\mathbf{c}})^H\right] 
    & = \sum_{l=0}^{L} \sum_{q=0}^{Q}\frac{\sigma_{c_{l,q}}^2 \sigma_{\mathbf{n}}^2}{\sigma_{\mathbf{n}}^2 + \sigma_{c_{l,q}}^2 P_\textrm{p}}. 
\end{split}
    \label{eq:MSEsimplified}
\end{equation}

We can then combine \eqref{eq:approximation} and \eqref{eq:MSEsimplified} to rewrite $\mathbf{R}_\mathbf{v}$ as
\begin{equation}
\begin{split}
    \mathbf{R}_\mathbf{v} 
    & \preceq \frac{P_\textrm{c}}{K_\textrm{c}} \mathbb{E}\left[\textrm{tr}(\mathbf{c}-\hat{\mathbf{c}})(\mathbf{c}-\hat{\mathbf{c}})^H\right] \mathbf{I}_{ R_\textrm{c}} + \sigma_{\mathbf{n}}^2 \mathbf{I}_{R_\textrm{c}} 
    % \\
    % & = 
    \left[\frac{P_\textrm{c}}{K_\textrm{c}}\sum_{l=0}^{L} \sum_{q=0}^{Q}\frac{\sigma_{c_{l,q}}^2 \sigma_{\mathbf{n}}^2}{\sigma_{\mathbf{n}}^2 + \sigma_{c_{l,q}}^2 P_\textrm{p}}  + \sigma_{\mathbf{n}}^2\right]\mathbf{I}_{ R_\textrm{c}}. \\
    \label{eq:Rvupperbound}
\end{split}
\end{equation}
Finally we can derive a looser lower bound on the capacity from \eqref{eq:Cgeq} as %given in \eqref{eq:Clower1}.
% \begin{figure*}[b]
% \hrulefill
\begin{equation}
\begin{split}
C
    & \geq \frac{1}{K}\mathbb{E}\left[  \log \det  \left(\mathbf{I}_{R_\textrm{c}}+\frac{P_\textrm{c}}{K_\textrm{c}}\mathbf{R}^{-1}_\mathbf{v}\hat{\Tilde{\mathbf{H}}}_\textrm{c}\hat{\Tilde{\mathbf{H}}}_\textrm{c}^H\right)\right] \\
    & \geq \frac{1}{K}\mathbb{E}\left[  \log \det  \left(\mathbf{I}_{R_\textrm{c}}+\frac{P_\textrm{c}}{K_\textrm{c}}\left[\frac{P_\textrm{c}}{K_\textrm{c}}\sum_{l=0}^{L} \sum_{q=0}^{Q}\frac{\sigma_{c_{l,q}}^2 \sigma_{\mathbf{n}}^2}{\sigma_{\mathbf{n}}^2 + \sigma_{c_{l,q}}^2 P_\textrm{p}}  + \sigma_{\mathbf{n}}^2\right]^{-1}\hat{\Tilde{\mathbf{H}}}_\textrm{c}\hat{\Tilde{\mathbf{H}}}_\textrm{c}^H\right)\right]  = \underline{C}\\
\end{split}
\label{eq:Clower1}
\end{equation}
% \vspace{-0.6cm}
% \end{figure*}
Now lastly, we rewrite $\hat{\Tilde{\mathbf{H}}}_\textrm{c}\hat{\Tilde{\mathbf{H}}}_\textrm{c}^H$. 

We can show that (see Appendix \ref{app:derivation2} for the derivation)
\begin{equation}
   \textrm{tr}\left( \mathbb{E}\left[\hat{\Tilde{\mathbf{H}}}_\textrm{c}\hat{\Tilde{\mathbf{H}}}_\textrm{c}^H\right]\right) \leq R_\textrm{c} \sum_{q=-Q/2}^{Q/2}\sum_{l=0}^L \sigma^2_{\hat{c}_{l,q}}.
\end{equation}
This motivates us to normalize the channel matrix $\hat{\Tilde{\mathbf{H}}}_\textrm{c}$ as,
\begin{equation}
    \hat{\Tilde{\mathbf{H}}}_\textrm{c} = \sqrt{R_\textrm{c} \sum_{q=-Q/2}^{Q/2}\sum_{l=0}^L \sigma^2_{\hat{c}_{l,q}} }
    \hat{\Tilde{\mathbf{H}}}_\textrm{c}',
\end{equation}
where $\hat{\Tilde{\mathbf{H}}}_\textrm{c}'$ is the normalized channel matrix.
We substitute this normalization and the expression obtained in \eqref{eq:Rvupperbound} in the lower bound on the capacity ($\underline{C}$ in \eqref{eq:Clower1}) and obtain,% \eqref{eq:Clower2}
%
% \begin{figure*}[b]
% \hrulefill
\begin{equation}
\begin{split}
    \underline{C}
    & = \frac{1}{K}\mathbb{E}\left[  \log \det  \left(\mathbf{I}_{R_\textrm{c}}+\frac{P_\textrm{c}}{K_\textrm{c}}\left[\frac{P_\textrm{c}}{K_\textrm{c}} \sum_{l=0}^{L} \sum_{q=-Q/2}^{Q/2} 
    \frac{\sigma_{c_{l,q}}^2 \sigma_{\mathbf{n}}^2}{\sigma_{\mathbf{n}}^2 + \sigma_{c_{l,q}}^2 P_\textrm{p}} + 
    \sigma_{\mathbf{n}}^2 \right]^{-1}R_\textrm{c} \sum_{q=-Q/2}^{Q/2}\sum_{l=0}^L \sigma^2_{\hat{c}_{l,q}}
    \hat{\Tilde{\mathbf{H}}}_\textrm{c}'\hat{\Tilde{\mathbf{H}}}_\textrm{c}'^H\right)\right] \\    
    & = \frac{1}{K}\mathbb{E}\left[  \log \det  \left(\mathbf{I}_{R_\textrm{c}}+\rho
    \hat{\Tilde{\mathbf{H}}}_\textrm{c}'\hat{\Tilde{\mathbf{H}}}_\textrm{c}'^H\right)\right].  \\   
\end{split}
\label{eq:Clower2}
\end{equation}
% \vspace{-0.6cm}
% \end{figure*}
%
where
\begin{equation*}
\begin{split}
    \rho 
    & = \frac{P_\textrm{c} R_\textrm{c}}{K_\textrm{c}}\left[\frac{P_\textrm{c}}{K_\textrm{c}} \sum_{l=0}^{L} \sum_{q=-Q/2}^{Q/2} 
    \frac{\sigma_{c_{l,q}}^2 \sigma_{\mathbf{n}}^2}{\sigma_{\mathbf{n}}^2 + \sigma_{c_{l,q}}^2 P_\textrm{p}} + 
    \sigma_{\mathbf{n}}^2 \right]^{-1} %\times \\    &\hspace{5cm}
    \sum_{q=-Q/2}^{Q/2}\sum_{l=0}^L \sigma^2_{\hat{c}_{l,q}}.
\end{split}
\end{equation*}

We can maximize $\rho$ with respect to $P_\textrm{c}$ and $P_\textrm{p}$ to optimize $\underline{C}$ in \eqref{eq:Clower2} (since normalized channel matrix is independent of the power distribution). Let $P$ be the total transmitter power, and set $P_\textrm{c} = \alpha P$ and $P_\textrm{p} = (1-\alpha)P$; thus, $\rho$ becomes a function of $\alpha$.
The optimal power distribution is given by
\begin{equation}
    \alpha^\ast = \arg\max_\alpha \rho.
\end{equation}

\section{Simulations}
In this section, we validate the theoretical findings and compare our framework with related work.
\subsection{Numerical validation}
\begin{figure*}[tb]
\centering
\begin{subfigure}{.33\textwidth}
  \centering
  \includegraphics[width=1\textwidth]{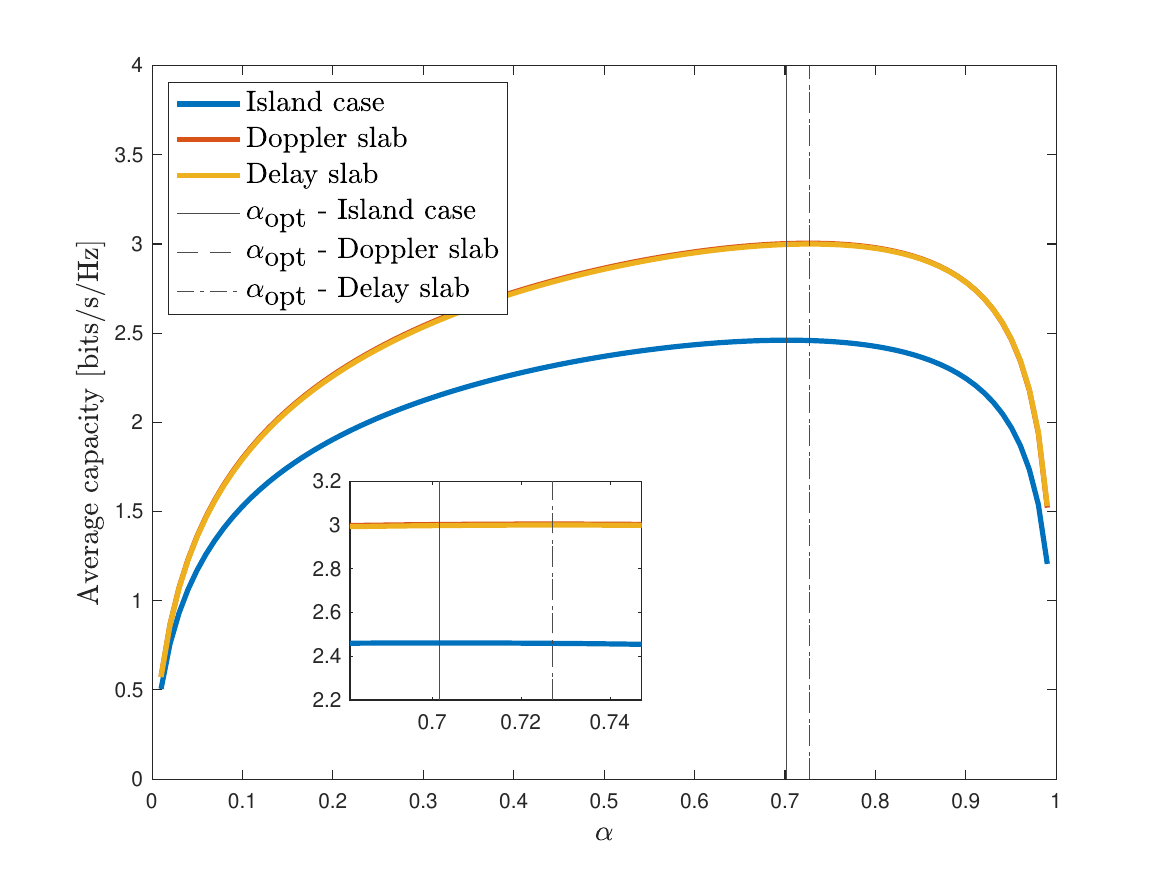}
  \caption{Channel 1: $Q = 6$, $L = 6$}
  \label{fig:AvCapCh1}
\end{subfigure}%
\begin{subfigure}{.33\textwidth}
  \centering
    \includegraphics[width=1\textwidth]{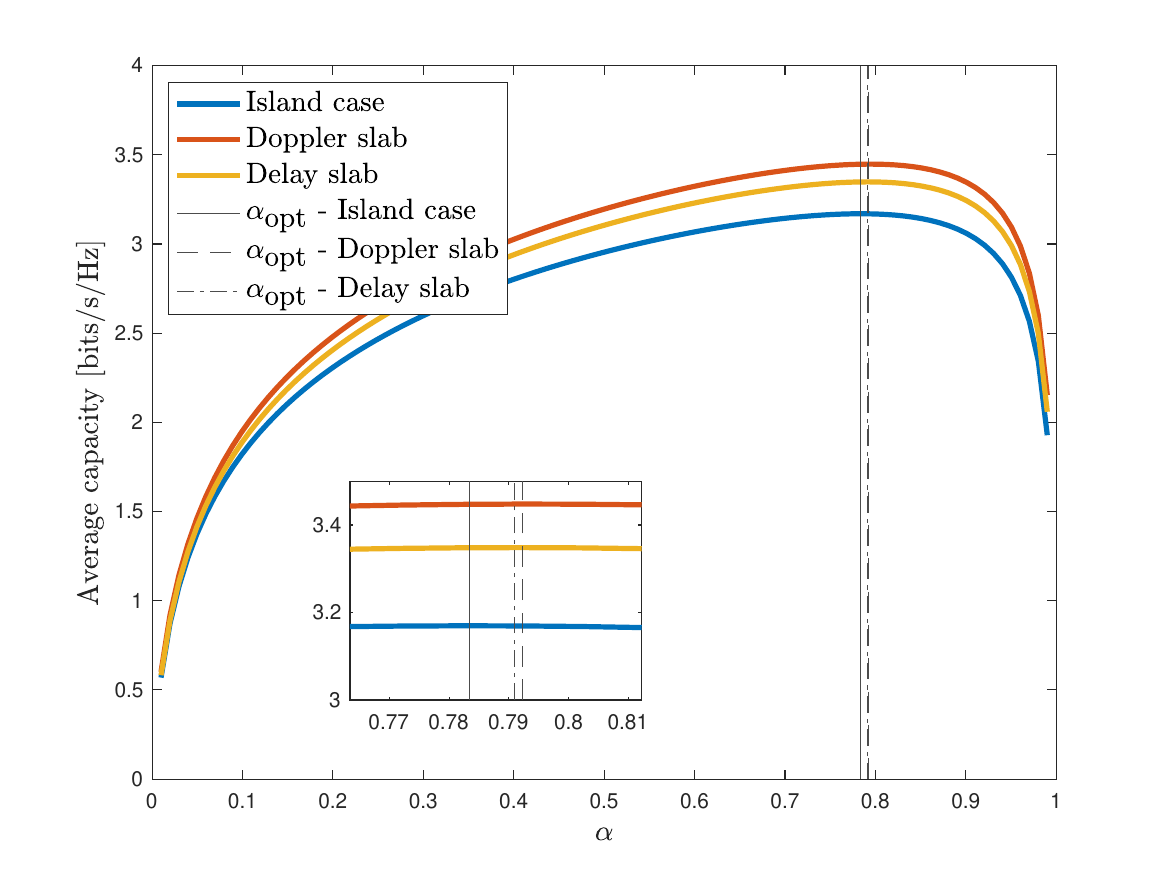}
  \caption{Channel 2:  $Q = 8$, $L = 2$}
  \label{fig:AvCapCh2}
\end{subfigure}
\begin{subfigure}{.33\textwidth}
  \centering
    \includegraphics[width=1\textwidth]{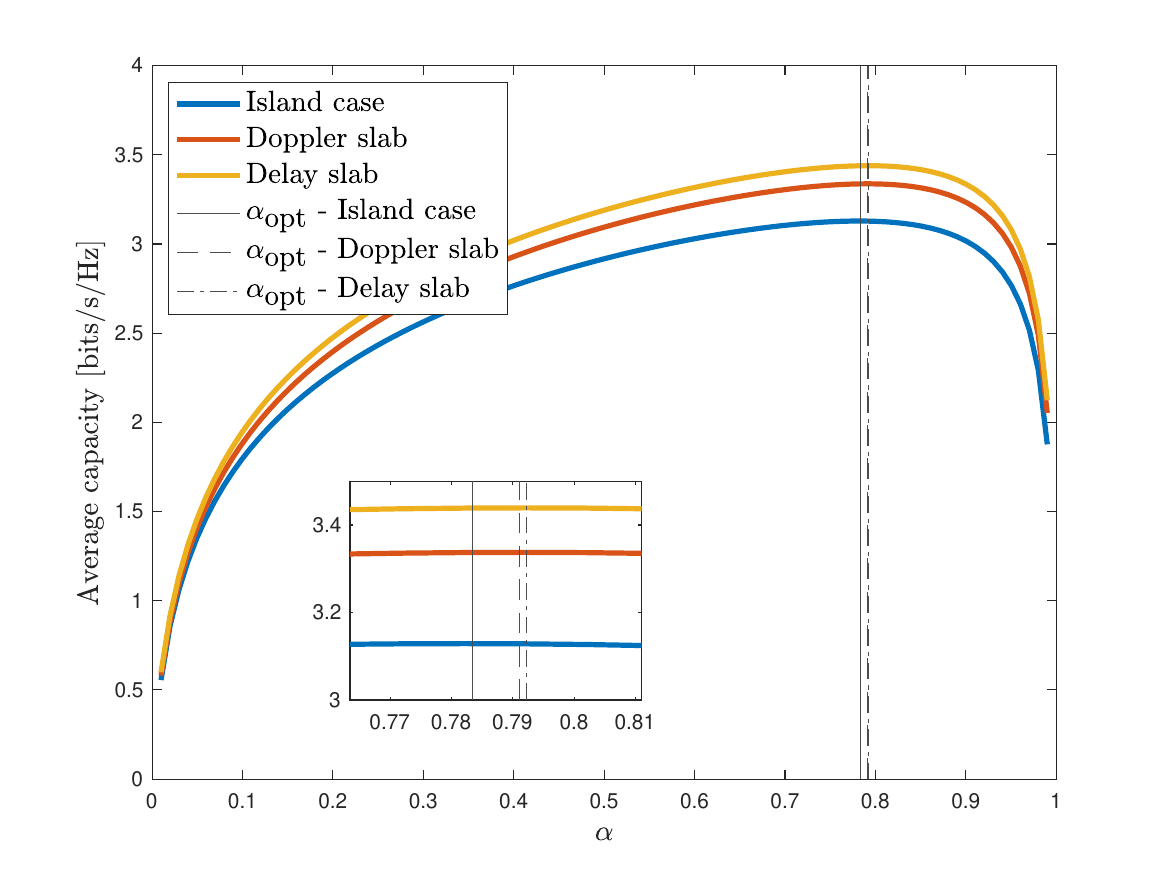}
  \caption{Channel 3:  $Q = 2$, $L = 8$}
  \label{fig:avCapCh3}
\end{subfigure}
\caption{Performance of the pilot allocation schemes with respect to the power distribution for three different channels}
\label{fig:avCapPerf}
\vspace{-0.2cm}
\end{figure*}

\noindent We simulate three baseband channels (i.e., frequency shifts between $0$ and $1$ and time delays of multiples of the sampling frequency) with different parameters, which are specified in Table~\ref{tab:paramsSim}. 
\begin{table*}[tp]
\begin{center}
\caption{Parameters for the simulations in Fig. \ref{fig:avCapPerf}.}
\label{tab:paramsSim}
\begin{tabular}{|l||c|c|c|}
\hline
Parameters & Channel $1$ & Channel $2$ & Channel $3$ \\ 
\hline
$K$ & $441$   & $441$   & $441$\\
$Q$ & $6$   & $8$   & $2$\\
$L$ & $6$   & $2$   & $8$\\
$\textrm{SNR}_\textrm{tx}$& $20$ dB& $20$ dB&$20$ dB \\
$\sigma_{c_{l,q}}^2$ & $1/((Q+1)(L+1))$, \: $\forall\{l,q\}$ & $1/((Q+1)(L+1))$, \: $\forall\{l,q\}$ & $1/((Q+1)(L+1))$, \: $\forall\{l,q\}$\\
Island case - $ \{N,M,\alpha_\textrm{opt}\}$    & $\{21,21,0.7015\}$        & $\{21,21,0.7834\}$    & $\{21,21,0.7834\}$ \\
Doppler slab - $ \{N,M,\alpha_\textrm{opt}\}$   & $\{7,63,0.7270\}$           & $\{9,49,0.7922\}$     & $\{3,147,0.7910\}$ \\
Delay slab - $ \{N,M,\alpha_\textrm{opt}\}$     & $\{63,7,0.7270\}$       & $\{147,3,0.7910\}$ & $\{49,9,0.7922\}$ \\
\hline
\end{tabular}
\end{center}
\vspace{-0.7cm}
\end{table*}
We call a channel `Doppler dominant' if $Q>L$ and `delay dominant' if $Q<L$. Thus, Channel $2$ and Channel $3$ are Doppler and delay dominant, respectively. We define the signal-to-noise ratio with respect to the transmitted signal $\mathbf{x}$, that is, 
\[
\textrm{SNR}_\textrm{tx} = \frac{\mathbf{x}^H\mathbf{x}}{\mathbf{n}^H\mathbf{n}} = \frac{P}{K\sigma_\mathbf{n}^2}.
\]
We set $P=1$, and change the value of $\sigma_\mathbf{n}^2$ according to the desired $\textrm{SNR}_\textrm{tx}$.
In all simulations, the (pilot and communication) symbols are uncoded QPSK symbols. The channel coefficients $c_{l,q}$ are realizations of a (independent) complex Gaussian process with zero mean and variance $1/((Q+1)(L+1))$. For all three channels we draw ten noise and channel realization, and calculate the average capacity.
The results are shown in Fig. \ref{fig:avCapPerf}. We can draw the following conclusions. First of all, the Doppler slab and delay slab, which alter the modulation parameters $N$ and $M$ according to the channel, exhibit higher capacity in all three channels. Moreover, we note that either the Doppler slab or delay slab performs the best, according to whether the channel is Doppler or delay dominant. This is to be expected, since the Doppler (delay) slab has the lowest pilot overhead for a Doppler (delay) dominant channel. 
Finally, we can see that all three pilot allocations benefit from the optimal power allocation. We see that, indeed, the maximum is reached at the power distribution $\alpha_\textrm{opt}$.

\subsection{Comparison with related work}
\noindent We provide a brief comparison between our framework and the study by \cite{raviteja2019embedded}. To ensure fairness, we align the parameters used in \cite{raviteja2019embedded} with those discussed in this paper.

Let $s_\textrm{p}$ denote a pilot symbol and let $s_\textrm{c}$ denote a communication symbol, then in \cite{raviteja2019embedded} the pilot and communication SNR were defined per \textit{symbol}, that is,
\begin{equation*}
    \textrm{SNR}_\textrm{p} = \frac{|s_\textrm{p}|^2}{\sigma_\mathbf{n}^2},\quad \textrm{SNR}_\textrm{c} = \frac{\mathbb{E}[|s_\textrm{c}|^2]}{\sigma_\mathbf{n}^2}.
\end{equation*}
Note that the relation to our SNR of the transmitted signal (thus pilot and communication signal together) is given by $\textrm{SNR}_\textrm{tx} =\frac{1}{K}\textrm{SNR}_\textrm{p}+\frac{K_\textrm{c}}{K}\textrm{SNR}_\textrm{c}$. We can relate the notion of SNR per \textit{symbol} of \cite{raviteja2019embedded} to our power distribution parameter,
\begin{equation*}
    \alpha = \frac{K_\textrm{c} \textrm{SNR}_\textrm{c}}{K_\textrm{c} \textrm{SNR}_\textrm{c} +\textrm{SNR}_\textrm{p}}.
\end{equation*}
Then, in Table \ref{tab:ravitejaresults} we list: the parameters that were used in \cite{raviteja2019embedded} (first four columns), ``our" parameters that follow from that (fifth to ninth column), and the parameters for the optimal pilot design (last four columns). We can see that the $\alpha$ that follows from \cite{raviteja2019embedded} differs a lot from the optimal distribution $\alpha^\ast$. In fact, we see that modifying the power distribution can lead to a substantial improvement in performance (compare seventh and ninth column). Furthermore, selecting the appropriate values for $M$ and $N$ can further enhance performance as we can see in the eleventh column.
\begin{table*}[tb]
\caption{Results of the comparison with \cite{raviteja2019embedded} for a simulated channel with $Q = 2$, $L = 6$.}
\centering
\begin{tabular}{|cccl|ccc|cc||cc|cc|}
\hline
\multicolumn{9}{|c||}{Island case} & \multicolumn{2}{c|}{Doppler slab} & \multicolumn{2}{c|}{Delay slab} \\
\multicolumn{9}{|c||}{$M = 128$, $N = 16$} & \multicolumn{2}{c|}{$M = 686$, $N = 3$} & \multicolumn{2}{c|}{$M = 7$, $N = 294$} \\
\hline
\multicolumn{4}{|c|}{Parameters from \cite{raviteja2019embedded}, cf Fig. 14} & \multicolumn{3}{c|}{Parameters resulting from \cite{raviteja2019embedded}}&\multicolumn{2}{c||}{Sub-optimal choice}&\multicolumn{2}{c|}{Sub-optimal choice:} & \multicolumn{2}{c|}{Optimal choice:}\\
\hline
$\textrm{SNR}_\textrm{p}$& $\textrm{SNR}_\textrm{c}$ & $\sigma^2$ & BER        & $\textrm{SNR}_\textrm{tx}$ & $\alpha$ & $\underline{C}(\alpha)$ & $\alpha^\ast$ & $\underline{C}(\alpha^\ast)$ & $\alpha^\ast$ & $\underline{C}(\alpha^\ast)$ & $\alpha^\ast$ & $\underline{C}(\alpha^\ast)$ \\
$[\SI{}{\decibel}]$ & $[\SI{}{\decibel}]$ & & & $[\SI{}{\decibel}]$ &  & $[\SI{}{\bit/\second/\hertz}]$&  & $[\SI{}{\bit/\second/\hertz}]$ &  & $[\SI{}{\bit/\second/\hertz}]$ &  & $[\SI{}{\bit/\second/\hertz}]$\\
\hline
$50$              & $20$              & $1$        & $\approx 1\cdot10^{-2}$  & $21.63$        & $0.6648$ & $3.9241$& $0.9064$ & $4.1396$ &$0.9072$& $4.1728$&$0.9072$ &$4.1774$\\
$60$              & $20$              & $1$        & $\approx 8.5\cdot10^{-3}$& $27.67$        & $0.1655$ & $4.0060$ & $0.9066$ & $5.4996$ &$0.9074$& $5.5495$&$0.9075$&$5.5582$\\
$50$              & $25$              & $1$        & $\approx 2\cdot10^{-3}$  & $25.50$        & $0.8625$ & $5.0351$ & $0.9066$ & $5.0482$ &$0.9073$& $5.0930$&$0.9074$&$5.1011$\\
$60$              & $25$              & $1$        & $\approx 1.5\cdot10^{-3}$& $29.00$        & $0.3854$ & $5.0897$ & $0.9066$ & $5.7523$ &$0.9074$& $5.8057$&$0.9075$&$5.8146$\\
\hline
\end{tabular}
\label{tab:ravitejaresults}
\vspace{-0.5cm}
\end{table*}
Note that an increase in $\textrm{SNR}_\textrm{tx}$ from $21.63$ dB to $27.67$ dB and from $25.50$ dB to $29.00$ dB is more than double the amount of power. It is important to highlight that this increase in power leads to only a minor Bit Error Rate (BER) improvement in \cite{raviteja2019embedded} as well as a minor capacity improvement as seen in the seventh column. By contrasting the optimal power allocation with the actual power distribution, the slight increase in BER becomes more understandable; the power increase is counteracted by the bad power distribution.

\section{Conclusions}
\noindent In this paper, we have aimed to contribute to the understanding of how to design pilot signals for OTFS.
We conducted an investigation into the literature on pilot design and established connections between the work on LTV channels, for both the OTFS modulation and OSDM modulation.

We have identified two minimum overhead pilot allocations for OTFS, that adjust $M$ or $N$ according to the channel parameters $L$ or $Q$, and show that these achieve the minimal MSE for channel estimation. In particular, the MMSE achieved is solely dependent on the total pilot power. 
We have also addressed the aspect of power distribution optimization with respect to the average capacity of the channel. Our results indicate that selecting an appropriate power distribution significantly enhances the (lower bound on the) average capacity of the communication system.

In summary, our research demonstrates that the careful selection of OTFS parameters, together with pilot design (the allocation and the power distribution) can lead to a significant improvement in average channel capacity.

% \section{Future Work}
% \noindent Subsequent research in this field could investigate further intricacies and practical limitations, such as constraints related to hardware capabilities. For instance, the suggested pilot configuration exhibits a high PAPR. In upcoming studies, we aim to consider the PAPR as a factor in the pilot signal design trade-off. 

\section{Acknowledgment}
\noindent This work was partly funded by the Netherlands Organisation for Applied Scientific Research (TNO) and the Netherlands Defence Academy (NLDA), reference no. TNO-10026587.

\appendices
\section{Relationship of BEM to delay-Doppler channel model}\label{app:ddchanisBEM}
\noindent The delay-Doppler channel assumes the channel consists of $P$ (narrowband) paths, each with a single delay and Doppler shift:
\begin{equation*}
\begin{split}
    h(t,\tau) 
    & = \sum_{p=1}^P h_\textrm{p} e^{j2\pi \nu_\textrm{p}(t-\tau_\textrm{p})}\delta(\tau-\tau_\textrm{p}). \\
\end{split}
\end{equation*}
Note that the assumption is that both the delay $\tau_i$ and Doppler shift $\nu_i$ are on a uniform grid, and are within the Nyquist-rate sampling time and frequency domain. Without loss of generality, one can set the points $(\tau_i,\nu_i)$ on a rectangular grid to size $(L+1)$ by $(Q+1)$. Let $p = q(L + 1) + l + 1$, where $q = 0,1,\dots,Q$ and $l = 0,1,\dots,L$. Note that $\tau_l = \tau_{l+L+1}$ and $\nu_{q(L+1)+1} = \nu_{q(L+1)+1+l}$, for $l = 0,\dots,L$. Then one can rewrite the time-varying impulse response as, 
\begin{equation*}
\begin{split}
    & h(t,\tau) 
    = \sum_{p=1}^P h_\textrm{p} e^{j2\pi \nu_\textrm{p}(t-\tau_\textrm{p})}\delta(\tau-\tau_\textrm{p}) \\
    & = \sum_{l=0}^L \sum_{q=-Q/2}^{Q/2} h_{q(L + 1) + l + 1} e^{j2\pi \nu_{q(L + 1) + l + 1}(t-\tau_{q(L + 1) + l + 1})} \times \\
    & \hspace{4cm}\quad\quad\quad\delta(\tau-\tau_{q(L + 1) + l + 1}) \\ 
    & = \sum_{l=0}^L \sum_{q=-Q/2}^{Q/2} h_{q,l} e^{j2\pi \nu_{q(L + 1) + 1}(t-\tau_{l + 1})}\delta(\tau-\tau_{l + 1}). \\ 
\end{split}
\end{equation*}
Note that $\nu_{q(L + 1) + 1}$ and $\tau_{l + 1}$ depend only on the indices $q$ and $l$, respectively. Therefore, without loss of generality, we can rewrite the channel as
\begin{equation*}
\begin{split}
    h(t,\tau) 
    & = \sum_{l=0}^L \sum_{q=-Q/2}^{Q/2} h_{q,l} e^{j2\pi \nu_{q}(t-\tau_{l})}\delta(\tau-\tau_{l}). \\ 
\end{split}
\end{equation*}
Its discrete counter part can then be written as
\begin{equation*}
\begin{split}
    & h(n,l)
    %&
    = \sum_{l'=0}^L \sum_{q=-Q/2}^{Q/2} h_{q,l'} e^{j2\pi \nu_{q}(n-\tau_{l'})}\delta(l-\tau_{l'})  \\
    & = \hspace{-0.2cm}\sum_{q=-Q/2}^{Q/2} h_{q,l} e^{j2\pi \nu_{q}(n-\tau_{l})} 
    %\\ &
    = \sum_{q=-Q/2}^{Q/2} h_{q,l}e^{-j2\pi \nu_{q}\tau_{l}} e^{j2\pi \nu_{q}n}.
\end{split}
\end{equation*}
It is now trivial to see that the models co-inside, since, w.l.o.g. we can set $c_{q,l}=h_{q,l'}e^{-j2\pi \nu_{q}\tau_{l'}}$. To conclude, the delay-Doppler channel is equivalent to the CE-BEM but written slightly differently.

\section{Derivation 1}\label{app:derivation}
\noindent We derive an upper bound as follows.
% \noindent See \eqref{eq:derivation1} at the bottom of the page.
% \begin{figure*}[b]
% \hrulefill
\begin{equation}
\begin{split}
    & \mathbb{E}[(\Tilde{\mathbf{H}}_\textrm{c}-\hat{\Tilde{\mathbf{H}}}_\textrm{c})(\Tilde{\mathbf{H}}_\textrm{c}-\hat{\Tilde{\mathbf{H}}}_\textrm{c})^H] \\
    & = \mathbb{E}\left[
    \left( \bm{\Psi}^H_\textrm{c}\left(\mathbf{H}_\textrm{DD} -\hat{\mathbf{H}}_\textrm{DD}\right)\bm{\Phi}_\textrm{c}\right)
    \left( \bm{\Psi}^H_\textrm{c}\left(\mathbf{H}_\textrm{DD} -\hat{\mathbf{H}}_\textrm{DD}\right)\bm{\Phi}_\textrm{c}\right)^H
    \right] \\
    & = \mathbb{E}\Bigg[
    \left(\sum_{q=-Q/2}^{Q/2}\sum_{l=0}^L ({c}_{l,q}-\hat{c}_{l,q})\bm{\Psi}^H_\textrm{c}(\mathbf{F}_N \otimes \mathbf{I}_M)(\bm{\Lambda}_q\mathbf{P}^l)(\mathbf{F}^H_N \otimes \mathbf{I}_M)\bm{\Phi}_\textrm{c}\right)
    \times \\
    & \hspace{5cm}
    \left(\sum_{q=-Q/2}^{Q/2}\sum_{l=0}^L ({c}_{l,q}-\hat{c}_{l,q})\bm{\Psi}^H_\textrm{c}(\mathbf{F}_N \otimes \mathbf{I}_M)(\bm{\Lambda}_q\mathbf{P}^l)(\mathbf{F}^H_N \otimes \mathbf{I}_M)\bm{\Phi}_\textrm{c}\right)^H
    \Bigg] \\
    & = \mathbb{E}\left[
    \sum_{q=-Q/2}^{Q/2}\sum_{l=0}^L |({c}_{l,q}-\hat{c}_{l,q})|^2
    \underbrace{
    \bm{\Psi}^H_\textrm{c}(\mathbf{F}_N \otimes \mathbf{I}_M)(\bm{\Lambda}_q\mathbf{P}^l)(\mathbf{F}^H_N \otimes \mathbf{I}_M)\bm{\Phi}_\textrm{c}
    \left(\bm{\Psi}^H_\textrm{c}(\mathbf{F}_N \otimes \mathbf{I}_M)(\bm{\Lambda}_q\mathbf{P}^l)(\mathbf{F}^H_N \otimes \mathbf{I}_M)\bm{\Phi}_\textrm{c}\right)^H
    }_{\textrm{Identity matrix with some zeros instead of ones}}
    \right] \\
    & \preceq \mathbb{E}\left[\textrm{tr}\left((\mathbf{c}-\hat{\mathbf{c}})(\mathbf{c}-\hat{\mathbf{c}})^H\right)\right]\mathbf{I}_{R_\textrm{c}}
\end{split}
\label{eq:derivation1}
\end{equation}
% \vspace{-0.6cm}
% \end{figure*}
\section{Derivation 2}\label{app:derivation2}
\noindent We can write, 
% \noindent We can write \eqref{eq:derivation2step2} (at the bottom of the next page)
%
% \begin{figure*}[b]
%     \hrulefill
\begin{align}
% \begin{split}
    &\mathbb{E}\left[\hat{\Tilde{\mathbf{H}}}_\textrm{c}\hat{\Tilde{\mathbf{H}}}_\textrm{c}^H\right] 
    = \mathbb{E}\left[
    \left( \bm{\Psi}^H_\textrm{c}\hat{\mathbf{H}}_\textrm{DD}\bm{\Phi}_\textrm{c}\right)
    \left( \bm{\Psi}^H_\textrm{c}\hat{\mathbf{H}}_\textrm{DD}\bm{\Phi}_\textrm{c}\right)^H
    \right] \notag\\
    & = \mathbb{E}\left[
    \left(\sum_{q=-Q/2}^{Q/2}\sum_{l=0}^L \hat{c}_{l,q}\bm{\Psi}^H_\textrm{c}(\mathbf{F}_N \otimes \mathbf{I}_M)(\bm{\Lambda}_q\mathbf{P}^l)(\mathbf{F}^H_N \otimes \mathbf{I}_M)\bm{\Phi}_\textrm{c}\right)
    \left(\sum_{q=-Q/2}^{Q/2}\sum_{l=0}^L \hat{c}_{l,q}\bm{\Psi}^H_\textrm{c}(\mathbf{F}_N \otimes \mathbf{I}_M)(\bm{\Lambda}_q\mathbf{P}^l)(\mathbf{F}^H_N \otimes \mathbf{I}_M)\bm{\Phi}_\textrm{c}\right)^H
    \right] \notag\\
    & = 
    % \mathbb{E}\left[
    \sum_{q=-Q/2}^{Q/2}\sum_{l=0}^L \sigma^2_{\hat{c}_{l,q}}
    \underbrace{
    \left(\bm{\Psi}^H_\textrm{c}(\mathbf{F}_N \otimes \mathbf{I}_M)(\bm{\Lambda}_q\mathbf{P}^l)(\mathbf{F}^H_N \otimes \mathbf{I}_M)\bm{\Phi}_\textrm{c}\right)
    \left(\bm{\Psi}^H_\textrm{c}(\mathbf{F}_N \otimes \mathbf{I}_M)(\bm{\Lambda}_q\mathbf{P}^l)(\mathbf{F}^H_N \otimes \mathbf{I}_M)\bm{\Phi}_\textrm{c}\right)^H
    }_{\textrm{Identity matrix with some diagonal entries zero instead of one}}
    % \right]
    \notag\\    & 
     \preceq  
     % \mathbb{E}\left[
    \sum_{q=-Q/2}^{Q/2}\sum_{l=0}^L \sigma^2_{\hat{c}_{l,q}}
    % \right]
    \mathbf{I}_{R_\textrm{c}},     \label{eq:derivation2step2}
% \end{split}
\end{align}
% \vspace{-0.6cm}
% \end{figure*}
such that
\begin{equation*}
   \textrm{tr}\left( \mathbb{E}\left[\hat{\Tilde{\mathbf{H}}}_\textrm{c}\hat{\Tilde{\mathbf{H}}}_\textrm{c}^H\right]\right) \leq R_\textrm{c} \sum_{q=-Q/2}^{Q/2}\sum_{l=0}^L \sigma^2_{\hat{c}_{l,q}}.
\end{equation*}
Note that $\sigma^2_{\hat{c}_{l,q}}$ is defined as in \eqref{eq:derivation2step1}.
% \begin{figure*}[b]
%     \hrulefill
\begin{align}
% \begin{split}
    \mathbb{E}[\hat{\mathbf{c}}\hat{\mathbf{c}}^H] 
    & = \mathbb{E}\left[
    \left(\sigma_{\mathbf{n}}^2\mathbf{R}^{-1}_{\mathbf{c}}+\mathbf{Z}^H\mathbf{Z} \right)^{-1} \mathbf{Z}^H\left(\mathbf{Z}\mathbf{c} + \mathbf{w}_\textrm{p}\right)
    \left(\mathbf{Z}\mathbf{c} + \mathbf{w}_\textrm{p}\right)^H\mathbf{Z}\left(\sigma_{\mathbf{n}}^2\mathbf{R}^{-1}_{\mathbf{c}}+\mathbf{Z}^H\mathbf{Z} \right)^{-1}
    \right] \notag\\
    & = 
    \left(\sigma_{\mathbf{n}}^2\mathbf{R}^{-1}_{\mathbf{c}}+\mathbf{Z}^H\mathbf{Z} \right)^{-1} 
    \mathbb{E}\left[
    \mathbf{Z}^H\left(\mathbf{Z}\mathbf{c}\mathbf{c}^H\mathbf{Z}^H + \mathbf{w}_\textrm{p}\mathbf{w}_\textrm{p}^H\right) \mathbf{Z}
    \right]
    \left(\sigma_{\mathbf{n}}^2\mathbf{R}^{-1}_{\mathbf{c}}+\mathbf{Z}^H\mathbf{Z} \right)^{-1}
    \notag\\ & 
    = 
    \left(\sigma_{\mathbf{n}}^2\mathbf{R}^{-1}_{\mathbf{c}}+\mathbf{Z}^H\mathbf{Z} \right)^{-1} 
    \left[
    \mathbf{Z}^H\mathbf{Z}\mathbf{R}_\mathbf{c}\mathbf{Z}^H\mathbf{Z} + \mathbf{Z}^H\mathbf{R}_{\mathbf{w}_\textrm{p}}\mathbf{Z}
    \right]
    \left(\sigma_{\mathbf{n}}^2\mathbf{R}^{-1}_{\mathbf{c}}+\mathbf{Z}^H\mathbf{Z} \right)^{-1}
     \notag\\
    & = 
    \frac{P^2_\textrm{p}\textrm{diag}\left([\sigma_{c_{0,0}}^2,\dots, \sigma_{c_{L,Q}}^2]\right) + P_\textrm{p}\sigma_{\mathbf{n}}^2\mathbf{I}_{(L+1)(Q+1)\times (L+1)(Q+1)}}{\left(\sigma_{\mathbf{n}}^2\textrm{diag}\left([\sigma_{c_{0,0}}^2,\dots, \sigma_{c_{L,Q}}^2]\right)^{-1}+P_\textrm{p}\mathbf{I}_{(L+1)(Q+1)\times (L+1)(Q+1)}\right)^{2} }
     \notag\\
    & = \textrm{diag}\left(\left[\frac{P_\textrm{p} \sigma_{c_{0,0}}^4}{\sigma_{\mathbf{n}}^2 + \sigma_{c_{0,0}}^2 P_\textrm{p}}, \dots,  \frac{P_\textrm{p} \sigma_{c_{L,Q}}^4}{\sigma_{\mathbf{n}}^2 + \sigma_{c_{L,Q}}^2 P_\textrm{p}}\right]\right) 
    %\\  &
    =  \textrm{diag}\left(\left[
    \sigma^2_{\hat{c}_{0,0}}, \dots, \sigma^2_{\hat{c}_{L,Q}}\right]\right) \label{eq:derivation2step1}
% \end{split}
\end{align}
% \vspace{-0.6cm}
% \end{figure*}

%%%%%%%%%%%%%%%%%%%%%%%%%%%%%%%%%%%%%%%%%%%%%%%%%%%%%%%%%%%%%%%%%
%%%%    BIBLIOGRAPHY    %%%%%%%%%%%%%%%%%%%%%%%%%%%%%%%%%%%%%%%%%
%%%%%%%%%%%%%%%%%%%%%%%%%%%%%%%%%%%%%%%%%%%%%%%%%%%%%%%%%%%%%%%%%
%\clearpage
\bibliographystyle{IEEEtran}
% \bibliography{references}
\bibliography{references}

%%%%%%%%%%%%%%%%%%%%%%%%%%%%%%%%%%%%%%%%%%%%%%%%%%%%%%%%%%%%%%%%%
%%%%    APPENDIX    %%%%%%%%%%%%%%%%%%%%%%%%%%%%%%%%%%%%%%%%%%%%%
%%%%%%%%%%%%%%%%%%%%%%%%%%%%%%%%%%%%%%%%%%%%%%%%%%%%%%%%%%%%%%%%%
%\newpage
%\appendices
%\input{IEEEBios}

\end{document}